\documentclass[twocolumn]{aastex631}

\usepackage{amsmath}
\usepackage{float}
\usepackage{natbib}
\usepackage{booktabs}
\usepackage{longtable}
\shorttitle{SNe 2018cni and 2020kyg}
\shortauthors{Mridweeka Singh et al.}

\begin{document}

\title{Observational properties of a bright type Iax SN 2018cni and a faint type Iax SN 2020kyg}

\correspondingauthor{Mridweeka Singh}
\email{mridweeka.singh@iiap.res.in, yashasvi04@gmail.com}

\author[0000-0001-6706-2749]{Mridweeka Singh}
\affiliation{Indian Institute of Astrophysics, Koramangala 2nd Block, Bangalore 560034, India}

\author[0000-0002-6688-0800]{Devendra. K. Sahu}
\affiliation{Indian Institute of Astrophysics, Koramangala 2nd Block, Bangalore 560034, India}

\author[0000-0001-6191-7160]{Raya Dastidar}
\affiliation{Millennium Institute of Astrophysics (MAS), Nuncio Monsenor Sòtero Sanz 100, Providencia, Santiago RM, Chile}
\affiliation{Instituto de Astrofísica, Universidad Andres Bello, Fernandez Concha 700, Las Condes, Santiago RM, Chile}

\author{Barnab\'as Barna}
\affiliation{Department of Experimental Physics, Institute of Physics, University of Szeged, H-6720 Szeged, D{\'o}m t{\'e}r 9, Hungary}

\author[0000-0003-1637-267X]{Kuntal Misra}
\affiliation{Aryabhatta Research Institute of observational sciencES, Manora Peak, Nainital, 263 001 India}

\author[0000-0002-3884-5637]{Anjasha Gangopadhyay}
\affiliation{Hiroshima Astrophysical Science Center, Hiroshima University, Higashi-Hiroshima, Japan}

\author{D. Andrew Howell}
\affiliation{Las Cumbres Observatory, 6740 Cortona Drive Suite 102, Goleta, CA, 93117-5575 USA}
\affiliation{Department of Physics, University of California, Santa Barbara, CA 93106-9530, USA}

\author[0000-0001-8738-6011]{Saurabh W. Jha}
\affiliation{Department of Physics and Astronomy, Rutgers, the State University of New Jersey, 136 Frelinghuysen Road, Piscataway, NJ 08854-8019, USA}

\author{Hyobin Im}
\affiliation{Korea Astronomy and Space Science Institute, 776 Daedeokdae-ro, Yuseong-gu, Daejeon 34055, Republic of Korea}
\affiliation{Korea University of Science and Technology (UST), 217 Gajeong-ro, Yuseong-gu, Daejeon 34113, Republic of Korea}

\author[0000-0002-5748-4558]{Kirsty Taggart}
\affiliation{Department of Astronomy and Astrophysics, University of California, Santa Cruz, CA 95064, USA}

\author{Jennifer Andrews}
\affiliation{Gemini Observatory, 670 North A‘ohoku Place, Hilo, HI 96720-2700, USA}

\author[0000-0002-1125-9187]{Daichi Hiramatsu}
\affiliation{Las Cumbres Observatory, 6740 Cortona Drive Suite 102, Goleta, CA, 93117-5575 USA}
\affiliation{Department of Physics, University of California, Santa Barbara, CA 93106-9530, USA}
\affiliation{Center for Astrophysics \textbar{} Harvard \& Smithsonian, 60 Garden Street, Cambridge, MA 02138-1516, USA}
\affiliation{The NSF AI Institute for Artificial Intelligence and Fundamental Interactions}

\author[0000-0002-0525-0872]{Rishabh Singh Teja}
\affiliation{Indian Institute of Astrophysics, Koramangala 2nd Block, Bangalore 560034, India}
\affiliation{Pondicherry University, R.V. Nagar, Kalapet, 605014, Puducherry, India}

\author{Craig Pellegrino}
\affiliation{Las Cumbres Observatory, 6740 Cortona Drive Suite 102, Goleta, CA, 93117-5575 USA}
\affiliation{Department of Physics, University of California, Santa Barbara, CA 93106-9530, USA}

\author{Ryan J. Foley}
\affiliation{Department of Astronomy and Astrophysics, University of California, Santa Cruz, CA 95064-1077, USA}

\author[0000-0001-9275-0287]{Arti Joshi}
\affiliation{Indian Institute of Astrophysics, Koramangala 2nd Block, Bangalore 560034, India}

\author{G. C. Anupama}
\affiliation{Indian Institute of Astrophysics, Koramangala 2nd Block, Bangalore 560034, India}

\author{K. Azalee Bostroem}
\affiliation{DiRAC Institute, Department of Astronomy, University of Washington, Box 351580, U.W., Seattle, WA 98195, USA}

\author[0000-0002-7472-1279]{Jamison Burke}
\affiliation{Las Cumbres Observatory, 6740 Cortona Drive Suite 102, Goleta, CA, 93117-5575 USA}
\affiliation{Department of Physics, University of California, Santa Barbara, CA 93106-9530, USA}

\author{Yssavo Camacho-Neves}
\affiliation{Department of Physics and Astronomy, Rutgers, the State University of New Jersey, 136 Frelinghuysen Road, Piscataway, NJ 08854-8019, USA}

\author{Anirban Dutta }
\affiliation{Indian Institute of Astrophysics, Koramangala 2nd Block, Bangalore 560034, India}
\affiliation{Pondicherry University, Chinna Kalapet, Kalapet, Puducherry 605014, India}

\author{Lindsey A. Kwok}
\affiliation{Department of Physics and Astronomy, Rutgers, the State University of New Jersey, 136 Frelinghuysen Road, Piscataway, NJ 08854-8019, USA}

\author{Curtis McCully}
\affiliation{Las Cumbres Observatory, 6740 Cortona Drive Suite 102, Goleta, CA, 93117-5575 USA}
\affiliation{Department of Physics, University of California, Santa Barbara, CA 93106-9530, USA}

\author{Yen-Chen Pan}
\affiliation{Graduate Institute of Astronomy, National Central University, 300 Zhongda Road, Zhongli, Taoyuan 32001, Taiwan}

\author{Matt Siebert}
\affiliation{Department of Astronomy and Astrophysics, University of California, Santa Cruz, CA 95064, USA}

\author[0000-0003-4524-6883]{Shubham Srivastav}
\affiliation{Astrophysics Research Centre, School of Mathematics and Physics, Queen’s University Belfast, Belfast BT7 1NN, UK}

\author{Tam\'as Szalai}
\affiliation{Department of Experimental Physics, Institute of Physics, University of Szeged, H-6720 Szeged, D{\'o}m t{\'e}r 9, Hungary}
\affiliation{ELKH-SZTE Stellar Astrophysics Research Group, H-6500 Baja, Szegedi {\'u}t, Kt. 766, Hungary}

\author[0000-0002-9486-818X]{Jonathan J. Swift}
\affiliation{Thacher School, 5025 Thacher Road, Ojai, CA 93023-8304, USA}

\author[0000-0001-7823-2627]{Grace Yang }
\affiliation{Thacher School, 5025 Thacher Road, Ojai, CA 93023-8304, USA}

\author[0000-0002-2093-6960]{Henry Zhou}
\affiliation{Thacher School, 5025 Thacher Road, Ojai, CA 93023-8304, USA}

\author[0000-0003-4522-9653]{Nico DiLullo}
\affiliation{Thacher School, 5025 Thacher Road, Ojai, CA 93023-8304, USA}

\author[0000-0002-3522-6312]{Jackson Scheer}
\affiliation{Thacher School, 5025 Thacher Road, Ojai, CA 93023-8304, USA}




\begin{abstract}
We present the optical photometric and spectroscopic analysis of two type Iax SNe 2018cni and 2020kyg. SN 2018cni is a bright type Iax SN (M$_{V,peak}$ = $-$17.81$\pm$0.21 mag) whereas SN 2020kyg (M$_{V,peak}$ = $-$14.52$\pm$0.21 mag) is a faint one. We derive $^{56}$Ni mass of 0.07 and 0.002 M${_\odot}$, ejecta mass of 0.48 and 0.14 M${_\odot}$ for SNe 2018cni and 2020kyg, respectively. A combined study of the bright and faint type Iax SNe in $R/r$- band reveals that the brighter objects tend to have a longer rise time. However, the correlation between the peak luminosity and decline rate shows that bright and faint type Iax SNe exhibit distinct behaviour. Comparison with standard deflagration models suggests that SN 2018cni is consistent with the deflagration of a CO white dwarf whereas the properties of SN 2020kyg can be better explained by the deflagration of a hybrid CONe white dwarf. The spectral features of both the SNe point to the presence of similar chemical species but with different mass fractions. Our spectral modelling indicates stratification at the outer layers and mixed inner ejecta for both the SNe.
\end{abstract}



\section{Introduction} \label{sec:intro}

Type Ia supernovae (SNe) are the final outcome of the thermonuclear explosion of a Carbon-Oxygen (CO) white dwarf in a binary system \citep{1960ApJ...132..565H,WANG2012122,doi:10.1142/S021827181630024X,2014ARA&A..52..107M,2019NatAs...3..706J}. As a group, type Ia SNe display great uniformity and are well known as members of one parameter family \citep{1993ApJ...413L.105P,1999AJ....118.1766P}. Despite the uniformity, diversity has also been noted from time to time based on the explosion scenario \citep{2017hsn..book..317T}. Amongst different subtypes of thermonuclear SNe, type Iax SNe emerge as low luminosity (M$_{V,peak}$ = $-$13 to $-$19 mag, \citealt{2017hsn..book..375J}) and less energetic members \citep{2013ApJ...767...57F}. They are also known as 2002cx-like SNe after its prototype SN 2002cx \citep{2003PASP..115..453L,2006AJ....131..527J}. \cite{2010ApJ...720..704M} suggested a correlation between peak luminosity, photospheric velocity at maximum and shape of the light curve of type Iax SNe, however, there are a few outliers such as SN 2009ku \citep{2011ApJ...731L..11N}, SN 2014ck \citep{2016MNRAS.459.1018T} etc. \cite{2011ApJ...731L..11N} suggested a scaling relation between the decline rate and ejecta velocity. A possible correlation between decline rate and absolute magnitude at maximum was also shown by \cite{2013ApJ...767...57F} and \cite{ps15csd2016}. The diversity in the observed properties such as peak brightness, light curve decline rates, and ejecta velocities indicates that type Iax SNe form a heterogeneous class. 

During the early phase, spectra of type Iax SNe exhibit similar spectral signatures as SN 1991T-like type Ia SNe \citep{2007PASP..119..360P,2010ApJ...708L..61F}. They show \ion{Fe}{2} and \ion{Fe}{3} lines during early spectral evolution and do not display any high velocity features \citep{2015MNRAS.451.1973S}. The brighter type Iax members such as SNe 2005hk \citep{2008ApJ...680..580S}, 2012Z \citep{stritzinger2015, 2015ApJ...806..191Y}, and 2020rea \citep{2022MNRAS.517.5617S} have weaker \ion{Si}{2} and less distinguished \ion{C}{2} features as compared to the fainter members of this class such as SNe 2008ha \citep{2009AJ....138..376F}, 2019gsc \citep{2020ApJ...892L..24S,2020MNRAS.496.1132T} and 2020kyg \citep{2022MNRAS.511.2708S}. Also, the W-shaped feature, attributed to \ion{S}{2} \citep{2006MNRAS.370..299H}, looks stronger in the spectra of the fainter type Iax than the brighter ones. 

The wide range of luminosity among members of this class presents a challenge in understanding them collectively in terms of a single physical scenario. For SN 2012Z, \cite{2014Natur.512...54M} reported the detection of a blue source coinciding with the SN location in the pre-explosion images. An accreting white dwarf with a Helium star in a binary system has been suggested as the plausible progenitor of SN 2012Z \citep{2014Natur.512...54M,2022ApJ...925..138M}. \cite{2012ApJ...761L..23J}, \cite{2013MNRAS.429.2287K} and \cite{2014MNRAS.438.1762F} proposed that the deflagration of a Chandrasekhar-mass white dwarf, leaving behind a remnant, could explain the low luminous nature of type Iax class. \cite{2014ApJ...792...29F} investigated the emission observed at the explosion site of SN 2008ha four years after the explosion and discussed the possibilities of its association with the remnant of the white dwarf. Further, the presence of a bound remnant is also supported by late time flattening found in the light curves of type Iax SNe \citep{2014ApJ...786..134M,2018MNRAS.474.2551S,10.1093/pasj/psab075,2022ApJ...925..138M}. A similar explosion of a hybrid Carbon Oxygen Neon (CONe) white dwarf \citep{2014ApJ...789L..45M,2015MNRAS.450.3045K} yields an outcome consistent with faint type Iax SNe. Given the variety of possible progenitor channels \citep{Pumo_2009, 2009Natur.459..674V,2010ApJ...719.1445M,2015MNRAS.447.2696D,2016A&A...589A..38B,2018ApJ...869..140K,2022MNRAS.510.3758B,2022A&A...658A.179L}, detailed analysis of a statistically complete sample of type Iax SNe is required to thoroughly understand the nature of this peculiar class of SNe.  

Bright and faint members of the type Iax class show diverse nature in terms of their luminosity, the synthesized mass of $^{56}$Ni, spectral features, etc. This paper presents an optical follow up of two type Iax SNe 2018cni and 2020kyg. SN 2018cni belongs to the bright type Iax SNe and SN 2020kyg is situated at the fainter end of the type Iax class. We intend to examine the heterogeneity or homogeneity in the two extremes of luminosity distribution of type Iax SNe. Section \ref{sec:Discovery, observation and data reduction} provides information about the discovery of the SNe, observations, and reduction of the data. Distance and extinction estimations are given in Section \ref{sec:Distance and extinction}. Section \ref{sec:Light curve properties} presents a detailed description of the photometric properties of the SNe and analytical modelling of the bolometric light curve. Section \ref{sec:Prime spectral features} discusses spectroscopic properties and spectral modelling. A comparative analysis followed by a brief summary is presented in Section \ref{sec:discussion_and_summary}. 

\begin{table*}
\caption{Optical photometric data of SN 2018cni}
\centering
\smallskip
\scriptsize
\begin{tabular}{c c c c c c c c  }
\hline \hline
Date    &   JD$^\dagger$   &   Phase$^\ddagger$ 	&   B       		     &      V           	 &  g                		& r   	        	 & i                   \\
        &                  &   (Days)           	& (mag)     		     & (mag)                     & (mag)             		&(mag) 			&(mag)            \\
\hline  

2018-06-19  & 288.64 & -1.34      &              18.46$\pm$ 0.01       &	    18.18$\pm$ 0.01    &      18.33$\pm$ 0.01    &      18.13$\pm$ 0.01  &   18.26$\pm$ 0.02    \\ 
2018-06-19  & 288.64 & -1.34      &              18.48$\pm$ 0.01       &            18.15$\pm$ 0.01    &      18.32$\pm$ 0.01    &      18.15$\pm$ 0.02  &   18.24$\pm$ 0.02   \\
2018-06-20  & 289.98 &  0.00      &              18.35$\pm$ 0.06       &            18.03$\pm$ 0.03    &      18.13$\pm$ 0.02    &      --     $\pm$ --  &   --   $\pm$ --   \\
2018-06-20  & 290.00 &  0.02      &              --     $\pm$ --       &            --     $\pm$ --    &      18.22$\pm$ 0.02    &      17.98$\pm$ 0.03  &   18.09$\pm$ 0.07   \\
2018-06-20  & 290.00 &  0.02      &              --     $\pm$ --       &            --     $\pm$ --    &      --     $\pm$ --    &      18.01$\pm$ 0.03  &   18.12$\pm$ 0.07   \\
2018-06-26  & 295.94 &  5.95      &              18.95$\pm$ 0.07       &            18.22$\pm$ 0.08    &      18.78$\pm$ 0.08    &      18.00$\pm$ 0.08  &   18.15$\pm$ 0.07   \\
2018-06-26  & 295.94 &  5.95      &              --      $\pm$ --      &            18.19$\pm$ 0.08    &      18.71$\pm$ 0.09    &      18.12$\pm$ 0.07  &   --   $\pm$ --    \\
2018-06-29  & 298.58 &  8.59      &              19.28$\pm$ 0.09       &            18.43$\pm$ 0.04    &      19.13$\pm$ 0.05    &      18.05$\pm$ 0.02  &   18.05$\pm$ 0.03   \\
2018-06-29  & 298.58 &  8.59      &              19.42$\pm$ 0.11       &            18.36$\pm$ 0.04    &      19.08$\pm$ 0.05    &      18.08$\pm$ 0.02  &   18.09$\pm$ 0.03   \\ 
2018-07-03  & 303.37 &  13.38     &              20.25$\pm$ 0.12       &            18.71$\pm$ 0.04    &      19.58$\pm$ 0.05    &      18.30$\pm$ 0.03  &   18.12$\pm$ 0.05   \\
2018-07-03  & 303.38 &  13.39     &              20.04$\pm$ 0.10       &            18.65$\pm$ 0.05    &      19.70$\pm$ 0.08    &      18.36$\pm$ 0.03  &   --   $\pm$    --        \\
2018-07-07  & 306.64 &  16.65     &              20.61$\pm$ 0.09       &            18.99$\pm$ 0.04    &      20.10$\pm$ 0.06    &      18.50$\pm$ 0.03  &   18.24$\pm$ 0.05   \\
2018-07-07  & 306.64 &  16.65     &              --      $\pm$ --      &            19.05$\pm$ 0.04    &      20.13$\pm$ 0.06    &      18.53$\pm$ 0.03  &   18.39$\pm$ 0.06   \\
2018-07-10  & 309.89 &  19.90     &              20.80$\pm$ 0.10       &            19.20$\pm$ 0.03    &      20.43$\pm$ 0.06    &      18.70$\pm$ 0.02  &   18.43$\pm$ 0.03   \\
2018-07-10  & 309.89 &  19.90     &              20.86$\pm$ 0.11       &            19.18$\pm$ 0.02    &      20.44$\pm$ 0.06    &      18.68$\pm$ 0.02  &   18.41$\pm$ 0.04   \\
2018-07-13  & 312.50 &  22.51     &              21.05$\pm$ 0.11       &            19.40$\pm$ 0.05    &      20.74$\pm$ 0.07    &      18.85$\pm$ 0.03  &   18.40$\pm$ 0.04   \\ 
2018-07-13  & 312.50 &  22.51     &              --     $\pm$ --       &            19.47$\pm$ 0.05    &      20.63$\pm$ 0.06    &      18.96$\pm$ 0.03  &   18.46$\pm$ 0.03   \\ 
2018-07-17  & 317.00 &  27.01     &              20.97$\pm$ 0.10       &            19.57$\pm$ 0.06    &      20.64$\pm$ 0.09    &      18.97$\pm$ 0.04  &   18.67$\pm$ 0.05   \\ 
2018-07-17  & 317.00 &  27.01     &              --     $\pm$ --       &            19.51$\pm$ 0.05    &      20.49$\pm$ 0.06    &      18.96$\pm$ 0.04  &   18.75$\pm$ 0.07   \\ 
2018-07-19  & 319.23 &  29.24     &              21.04$\pm$ 0.29       &            19.63$\pm$ 0.14    &      20.69$\pm$ 0.24    &      19.15$\pm$ 0.12  &   18.84$\pm$ 0.11   \\ 
2018-07-19  & 319.23 &  29.24     &              --     $\pm$ --       &            --     $\pm$ --    &      20.71$\pm$ 0.29    &      19.08$\pm$ 0.11  &   18.86$\pm$ 0.10   \\ 
2018-07-24  & 323.99 &  34.00     &              --     $\pm$ --       & 	      --   $\pm$    -- &      20.56$\pm$ 0.11    &      --     $\pm$ --  &   18.88$\pm$ 0.08   \\ 
2018-07-24  & 323.99 &  34.00     &              --     $\pm$ --       & 	      --   $\pm$    -- &      --     $\pm$ --    &      --     $\pm$ --  &   18.83$\pm$ 0.09   \\ 
2018-07-24  & 324.00 &  34.01     &              --     $\pm$ --       &               --  $\pm$    -- &      --     $\pm$ --    &      19.21$\pm$ 0.08  &   18.88$\pm$ 0.08   \\ 
2018-07-24  & 324.00 &  34.01     &              --     $\pm$ --       &               --  $\pm$    -- &      --     $\pm$ --    &      19.14$\pm$ 0.08  &   18.83$\pm$ 0.09   \\ 
2018-07-25  & 325.38 &  35.39     &              --     $\pm$ --       &            19.93$\pm$ 0.27    &      --     $\pm$ --    &      --     $\pm$ --  &   --   $\pm$ --       \\
2018-07-27  & 326.88 &  36.89     &              21.08  $\pm$ 0.19     & 	    19.76$\pm$ 0.06    &      --     $\pm$ --  	 &      --     $\pm$ --  &   --   $\pm$ --     \\
2018-07-27  & 326.89 &  36.90     &              --     $\pm$ --       &            19.81$\pm$ 0.09    &      --     $\pm$ --    &      --     $\pm$ --  &   --   $\pm$ --    \\
2018-07-27  & 327.38 &  37.39     &              --     $\pm$ --       &            --     $\pm$ --    &      21.03$\pm$ 0.14    &      19.26$\pm$ 0.07  &   18.85$\pm$ 0.06    \\
2018-07-27  & 327.39 &  37.40     &              --     $\pm$ --       &            --     $\pm$ --    &      20.97$\pm$ 0.15    &      19.39$\pm$ 0.10  &   --   $\pm$ --   \\
2018-07-29  & 328.57 &  38.58     &              21.02$\pm$ 0.43       &            --     $\pm$ --    &      --     $\pm$ --    &      --     $\pm$ --  &   --   $\pm$ --   \\ 
2018-07-29  & 328.57 &  38.58     &              --     $\pm$ --       &            19.86$\pm$ 0.28    &      --     $\pm$ --    &      --     $\pm$ --  &   --   $\pm$ --    \\
2018-07-29  & 329.46 &  39.47     &              --     $\pm$ --       &            --     $\pm$ --    &      20.97$\pm$ 0.14    &      19.52$\pm$ 0.06  &   19.26$\pm$ 0.09   \\ 
2018-07-29  & 329.46 &  39.47     &              --     $\pm$ --       &            --     $\pm$ --    &      --     $\pm$ --    &      19.50$\pm$ 0.07  &   19.29$\pm$ 0.10    \\
2018-08-04  & 334.97 &  44.98     &              21.28$\pm$ 0.20       &            19.91$\pm$ 0.07    &      20.94$\pm$ 0.13    &      19.48$\pm$ 0.06  &   19.27$\pm$ 0.11   \\
2018-08-04  & 334.97 &  44.98     &              21.26$\pm$ 0.17       &            19.87$\pm$ 0.06    &      20.98$\pm$ 0.15    &      19.35$\pm$ 0.07  &   19.29$\pm$ 0.10  \\
2018-08-10  & 340.56 &  50.57     &              21.57$\pm$ 0.18       &            20.11$\pm$ 0.08    &      20.95$\pm$ 0.13    &      19.72$\pm$ 0.06  &   19.39$\pm$ 0.06  \\ 
2018-08-10  & 340.57 &  50.58     &              --     $\pm$ --       &            20.12$\pm$ 0.08    &      21.02$\pm$ 0.12    &      19.71$\pm$ 0.05  &   19.56$\pm$ 0.09  \\
2018-08-20  & 350.54 &  60.55     &              --     $\pm$ --       &            20.11$\pm$ 0.09    &      20.85$\pm$ 0.11    &      19.83$\pm$ 0.09  &   19.77$\pm$ 0.13 \\ 
2018-08-22  & 352.51 &  62.52     &              21.43$\pm$ 0.48       &            20.32$\pm$ 0.19    &      20.78$\pm$ 0.30    &      19.79$\pm$ 0.12  &   19.77$\pm$ 0.18  \\
2018-08-22  & 352.52 &  62.53     &              --     $\pm$  --      & 	    --     $\pm$ --    &      20.75$\pm$ 0.26    &      --     $\pm$ --  &   --   $\pm$ --   \\
2018-08-31  & 362.22 &  72.23     &              --     $\pm$  --      & 	    20.30$\pm$ 0.11    &      21.31$\pm$ 0.14    &      20.25$\pm$ 0.10  &   19.79$\pm$ 0.17  \\

\hline    
\end{tabular}
\newline
$^\dagger$ JD 2,458,000+ ,
$^\ddagger$ Phase has been calculated with respect to B$_{max}$ =2458289.99  
\label{tab:photometric_observational_log_2018cni}                                                        
\end{table*}

\section{Discovery, observation, and data reduction}
\label{sec:Discovery, observation and data reduction}

\begin{figure*}
\centering
\begin{minipage}{.45\textwidth}
  \centering
  \includegraphics[width=\linewidth]{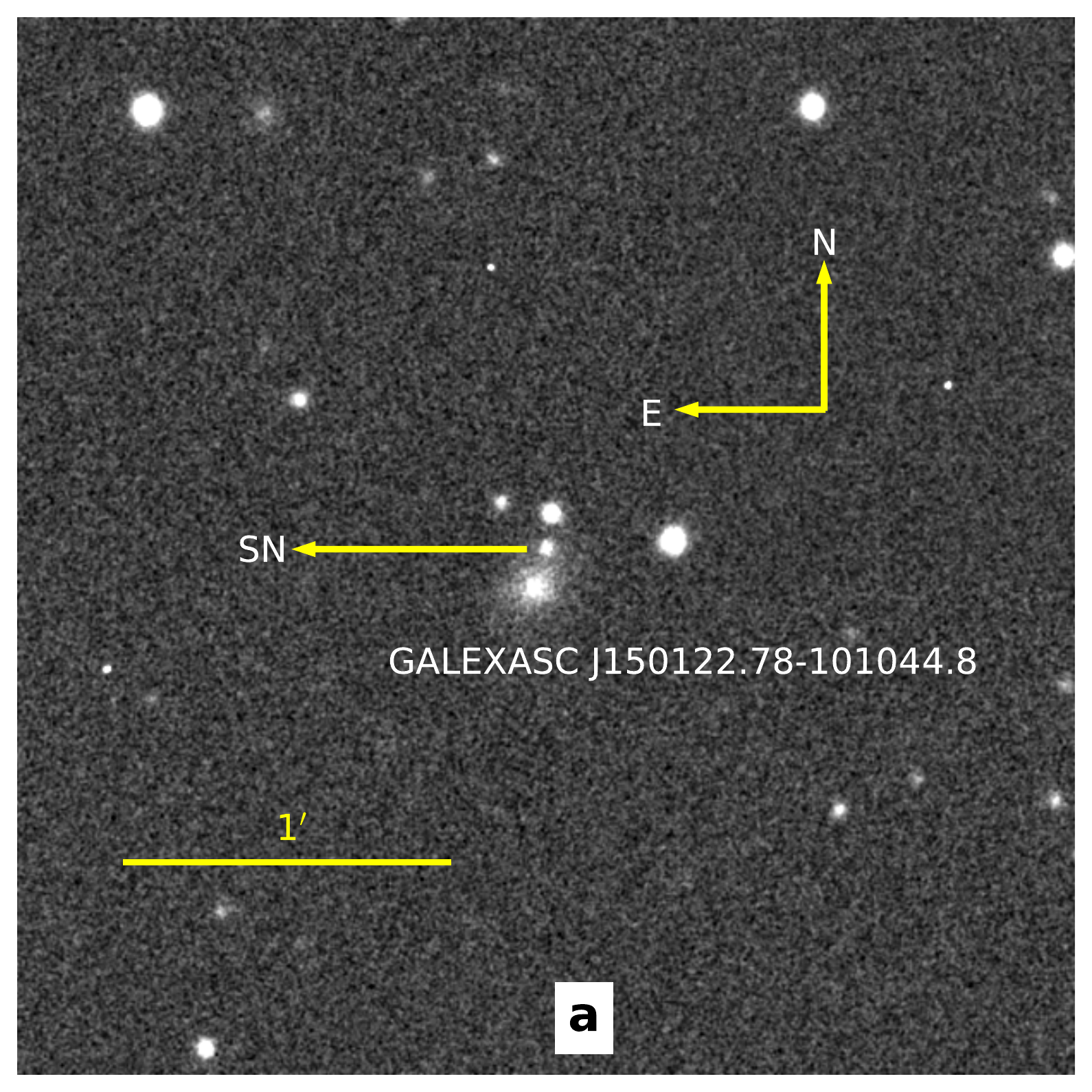}
  \label{fig:ds9_2018cni}
\end{minipage}%
\begin{minipage}{.45\textwidth}
  \centering
  \includegraphics[width=\linewidth]{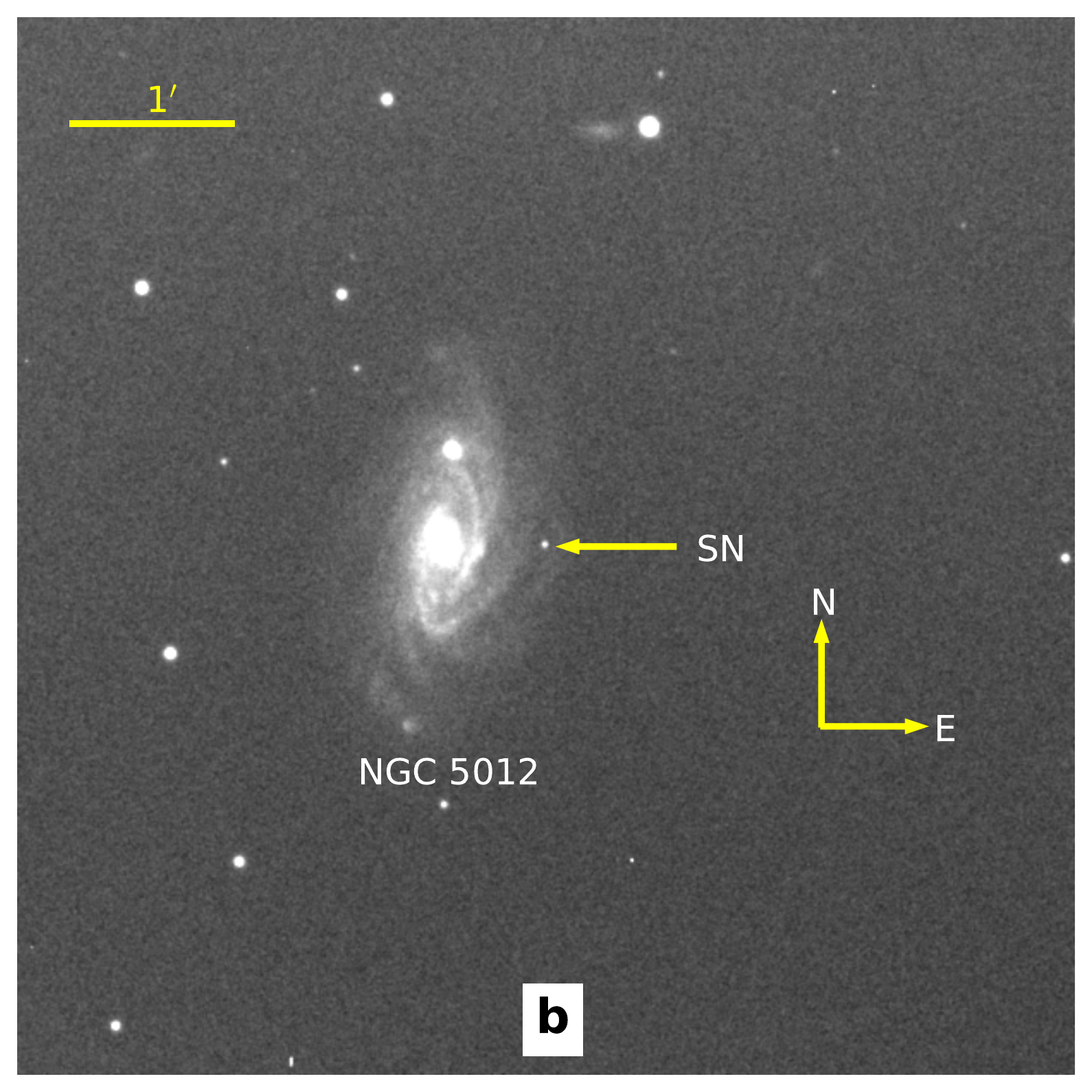}
  \label{fig:ds9_2020kyg}
\end{minipage}
\caption{(a) Location of SN 2018cni in its host galaxy. This image is acquired in \textit{V} band with  1-m LCO telescope on July 17, 2018. (b) SN 2020kyg and its host galaxy NGC 5012. This image is taken in \textit{V} band with  1-m LCO telescope on June 27, 2020.}
\end{figure*}

SN 2018cni (also known as PTSS-18fdb, ATLAS18qql and Gaia18bpu) was detected by PMO-Tsinghua Supernova Survey\footnote{PTSS, http://www.cneost.org/ptss/} \citep{2018TNSTR.820....1T} on June 13, 2018. The location of the SN is at 1$''$ east and 5$''$ south of the host galaxy GALEXASC J150122.78-101044.8. It was classified as a type Iax SN by \cite{2018ATel11728....1Z,2018TNSCR.846....1Z} using a spectrum obtained on June 17, 2018. This spectrum was found to match well with SN 2005hk, 3 days before maximum light. The location of SN 2018cni in its host galaxy is presented in Figure \ref{fig:ds9_2018cni}. 


\begin{table*}
\caption{Optical photometric data of SN 2020kyg}
\centering
\smallskip
\scriptsize
\begin{tabular}{c c c c c c c c  }
\hline \hline
Date    &   JD$^\dagger$      &   Phase$^\ddagger$   &   B           &      V                &  g                     & r   	                 & i                   \\
        &                     &   (Days)             &  (mag)        & (mag)                 & (mag)                  &(mag) 		         &(mag)        \\
\hline \hline
2020-05-27    &     996.88    &  -2.05     & 18.54$\pm$ 0.22  &  18.28$\pm$ 0.14  &  18.38$\pm$ 0.04   &       18.34$\pm$ 0.03  &	     18.55$\pm$ 0.03  \\  
2020-05-29    &     998.94    &   0.00     & 18.36$\pm$ 0.06  &  18.28$\pm$ 0.05  &  18.37$\pm$ 0.01   &       18.35$\pm$ 0.02  &        18.48$\pm$ 0.03   \\ 
2020-05-31    &     1000.92   &   1.98     & 18.56$\pm$ 0.06  &  18.36$\pm$ 0.06  &  18.48$\pm$ 0.02   &       18.31$\pm$ 0.03  &        18.38$\pm$ 0.04    \\
2020-06-01    &     1002.34   &   3.40     & 18.83$\pm$ 0.10  &  18.42$\pm$ 0.09  &  18.74$\pm$ 0.06   &       18.36$\pm$ 0.05  &        18.42$\pm$ 0.07  \\  
2020-06-05    &     1006.28   &   7.34     & 19.91$\pm$ 0.22  &  18.63$\pm$ 0.08  &  19.20$\pm$ 0.06   &       18.47$\pm$ 0.04  &        18.48$\pm$ 0.04   \\
2020-06-09    &     1010.21   &   11.27    & 20.69$\pm$ 0.11  &  19.17$\pm$ 0.06  &  20.07$\pm$ 0.05   &       18.76$\pm$ 0.02  &        18.65$\pm$ 0.03   \\
2020-06-15    &     1016.28   &   17.34    & 21.02$\pm$ 0.10  &  19.63$\pm$ 0.06  &  --   $\pm$ --     &       19.19$\pm$ 0.02  &        19.05$\pm$ 0.04   \\
2020-06-15    &     1016.28   &   17.34    & 21.14$\pm$ 0.12  &  19.54$\pm$ 0.06  &  --   $\pm$ --     &       19.24$\pm$ 0.02  &        19.12$\pm$ 0.04   \\
2020-06-15    &     1016.80   &   17.86    & --     $\pm$ --  &  19.59$\pm$ 0.06  &  --   $\pm$ --     &       --     $\pm$ --  &	 --    $\pm$  --     \\
2020-06-16    &     1016.91   &   17.97    & --     $\pm$ --  &  19.56$\pm$ 0.10  &  20.28$\pm$ 0.03   &       19.13$\pm$ 0.01  &	 18.93$\pm$ 0.01  \\
2020-06-19    &     1019.83   &   20.89    & 20.76$\pm$ 0.22  &  20.04$\pm$ 0.15  &  --   $\pm$ --   &       --     $\pm$ --  &	 --    $\pm$  --    \\
2020-06-22    &     1023.21   &   24.27    & 21.39$\pm$ 0.14  &  20.09$\pm$ 0.07  &  --   $\pm$ --   &       19.67$\pm$ 0.03  &	 19.37$\pm$ 0.04  \\
2020-06-22    &     1023.22   &   24.28    & --     $\pm$ --  &  20.03$\pm$ 0.07  &  --    $\pm$ --   &       19.57$\pm$ 0.03  &        19.43$\pm$ 0.05   \\
2020-06-27    &     1027.85   &   28.91    & 21.59$\pm$ 0.37  &  20.38$\pm$ 0.09  &  21.24$\pm$ 0.11   &       19.91$\pm$ 0.06  &        19.61$\pm$ 0.05   \\
2020-06-27    &     1027.85   &   28.91    & 21.64$\pm$ 0.32  &  20.11$\pm$ 0.09  &  21.81$\pm$ 0.20   &       19.84$\pm$ 0.04  &        19.81$\pm$ 0.07  \\
2020-07-01    &     1032.21   &   33.27    & --     $\pm$ --  &  20.43$\pm$ 0.17  &  21.26$\pm$ 0.20   &       20.09$\pm$ 0.11  &        19.74$\pm$ 0.10  \\
2020-07-01    &     1032.21   &   33.27    & --     $\pm$ --  &  20.62$\pm$ 0.22  &  21.75$\pm$ 0.43   &       19.94$\pm$ 0.07  &        --     $\pm$ --      \\
2020-07-06    &     1037.22   &   38.28    & --     $\pm$ --  &  20.67$\pm$ 0.17  &  21.62$\pm$ 0.31   &       20.40$\pm$ 0.12  &	 20.00$\pm$ 0.13  \\
2020-07-06    &     1037.22   &   38.28    & --     $\pm$ --  &  20.89$\pm$ 0.24  &  --     $\pm$ --   &       20.03$\pm$ 0.07  &        --     $\pm$ --     \\
2020-07-12    &     1043.21   &   44.27    & --     $\pm$ --  &  21.66$\pm$ 0.36  &  --     $\pm$ --   &       20.25$\pm$ 0.33  &	 --     $\pm$  --	   \\
2020-07-15    &     1045.87   &   46.93    & --     $\pm$ --  &  21.65$\pm$ 0.23  &  --     $\pm$ --   &       21.59$\pm$ 0.24  &	 20.11$\pm$ 0.13   \\  	
2020-07-22    &     1053.20   &   54.26    & --     $\pm$ --  &  21.32$\pm$ 0.24  &  --     $\pm$ --   &       20.96$\pm$ 0.14  &        20.54$\pm$ 0.11   \\ 
              
\hline    
\end{tabular}
\newline
$^\dagger$ JD 2,458,000+ ,
$^\ddagger$ Phase has been calculated with respect to B$_{max}$ = 2458998.94
\label{tab:photometric_observational_log_2020kyg}                                                        
\end{table*}

SN 2020kyg was spotted by Asteroid Terrestrial-impact Last Alert System \citep[ATLAS,][]{2018PASP..130f4505T} at 0.0$''$ north and 35.6$''$ east to the center of the host galaxy NGC 5012 on May 24, 2020 \citep{2020TNSTR1539....1T,2020TNSAN.113....1S}. ATLAS20nuc and AT2020kyg are other aliases of this source. A non-detection was reported on May 20, 2020 (MJD = 58989.36, \citealt{2020TNSAN.113....1S}). \cite{2020ATel13761....1O} and \cite{2020TNSCR1559....1H} classified it as a type Iax SN before maximum. Figure \ref{fig:ds9_2020kyg} shows the position of SN 2020kyg in the host galaxy NGC 5012.

Optical photometric observations of SNe 2018cni and 2020kyg were initiated $\sim$ 6 and 3 days from discovery with the 1-m telescopes of the Las Cumbres Observatory \citep[LCO,][]{2013PASP..125.1031B} under the Global Supernova Project (GSP) and continued up to $\sim$80 and 60 days, respectively. 
The photometry was performed using the \texttt{lcogtsnpipe} pipeline  \citep{2016MNRAS.459.3939V}. The $BV$ magnitudes are calibrated to the Vega system while the $gri$ magnitudes are calibrated to the AB system. SN 2018cni exploded in the proximity of the host galaxy, hence template subtraction was performed in order to eliminate the host galaxy contamination. The template was observed in $BVgri$ bands using the 1-m LCO telescope on August 6, 2019, which is more than a year after the discovery. The image subtraction was performed within the \texttt{lcogtsnpipe} using PyZOGY \citep{2017zndo...1043973G}. The optical photometry of SNe 2018cni and 2020kyg are presented in Tables \ref{tab:photometric_observational_log_2018cni} and \ref{tab:photometric_observational_log_2020kyg}, respectively. 

Spectroscopic observations of SNe 2018cni and 2020kyg were initiated $\sim$ 5 days and $\sim$ 1 day after the discovery, respectively. Four spectra of SN 2018cni and three spectra of SN 2020kyg were taken with FLOYDS spectrograph on the 2-m Faulkes Telescope North \citep[FTN,][]{2013PASP..125.1031B}. One spectrum of SN 2018cni was acquired with the Blue Channel (BC) spectrograph on 6.5-m MMT, and two spectra of SN 2018cni were obtained with Robert Stobie Spectrograph (RSS) mounted on 9.2-m Southern African Large Telescope (SALT). Two spectra of SN 2018cni were obtained using the dual-beam Kast spectrograph \citep{KAST} on the Lick Shane 3-m telescope. The 300/7500 grating on the red side and the 452/3306 grism on the blue side were used, providing continuous coverage between 3500–10400 \AA. The slit was aligned along the parallactic angle \citep{Filippenko1982}. Two spectra of SN 2020kyg were taken with Gemini Multi-Object Spectrograph (GMOS-N) using B600 and R400 gratings on Gemini North. Spectral reduction of all spectra taken with the FLOYDS spectrograph was done using the \texttt{floydsspec}\footnote{https://www.authorea.com/users/598/articles/6566} pipeline. The spectrum of SN 2018cni observed with MMT was reduced using standard tasks described in \citet{2019ApJ...885...43A}. The two spectra from the Kast spectrograph were reduced in a standard manner using tools in custom spectroscopy pipeline\footnote{\url{https://github.com/msiebert1/UCSC\_spectral\_pipeline}} \citep{Siebert2019}. The two spectra of SN 2018cni obtained from RSS were reduced using a custom pipeline which contains {\sc PYSALT} package and standard PYRAF spectral reduction tasks \citep{2010SPIE.7737E..25C}. The Gemini spectra of SN 2020kyg were reduced using the standard Gemini IRAF package. The log of spectroscopic observations of both the SNe are provided in Tables \ref{tab:spectroscopic_observations_cni} and \ref{tab:spectroscopic_observations_20kyg}, respectively. 

\begin{table*}
\caption{Log of spectroscopic observations for SN 2018cni}
\centering
\smallskip
\begin{tabular}{c c c c c  c}
\hline \hline
Date     & JD$^\dagger$     & Phase$^\ddagger$             & Spectral Range  & Resolution      & Telescope/Instrument       \\
          &                  &(Days)                             & (\AA)           &                 &                 \\
\hline
2018-06-18  & 287.79     & -2.2                 & 3300-11000   & 400-700   	   & FTN/FLOYDS  \\
2018-06-19  & 288.76     & -1.2                 & 3500-10400   &  400  	       &  SHANE/Kast \\
2018-06-29  & 298.81     & 8.8              	   & 3300-11000   & 400-700        & FTN/FLOYDS  \\
2018-07-05  & 304.14     & 14.2                 & 3900-6800    & 3900           & MMT/BCH  \\
2018-07-11  & 311.35     & 21.4                 & 3500-9300    & 1000           & SALT/RSS  \\
2018-07-20  & 319.72     & 29.7                 & 5620-10400   &  400           & SHANE/Kast  \\
2018-07-24  & 323.99     & 34.0                 & 3200-11000   & 400-700        & FTN/FLOYDS  \\
2018-07-26  & 325.98     & 35.9                 & 3200-11000   & 400-700        & FTN/FLOYDS  \\
2018-08-14  & 345.26     & 55.3             	   & 3500-9300    & 1000           & SALT/RSS  \\

\hline                                   
\end{tabular}
\newline
$^\dagger$2,458,000+
$^\ddagger$ Phase has been calculated with respect to $B$$_{max}$ = 2458289.99
\label{tab:spectroscopic_observations_cni}      
\end{table*}

\begin{table*}
\caption{Log of spectroscopic observations for SN 2020kyg}
\centering
\smallskip
\begin{tabular}{c c c c c c}
\hline \hline
Date     & JD$^\dagger$     & Phase$^\ddagger$    & Spectral Range  & Resolution      & Telescope/Instrument       \\
          &                   &(Days)                                & (\AA)           &                 &                 \\
\hline
2020-05-25  & 994.93    & -4.0  & 3300-11000   & 400-700	 &    FTN/FLOYDS  \\
2020-05-26  & 995.80    & -3.1  & 3200-10000   & 400-700     &    FTN/FLOYDS \\
2020-06-04  & 1004.85   &  5.9  & 3200-10000   & 400-700     &    FTN/FLOYDS  \\
2020-06-14  & 1014.81   &  15.8  & 4000-9000   & 1688,1900   &    Gemini/GMOS-N \\
2020-07-03  & 1033.83   &  34.9  & 4000-9000   & 1688,1900   &    Gemini/GMOS-N  \\

\hline                                   
\end{tabular}
\newline
$^\dagger$ 2,458,000+
$^\ddagger$ Phase is from B$_{max}$= 2458998.94
\label{tab:spectroscopic_observations_20kyg}      
\end{table*}

\begin{table*}
\caption{Parameters of SNe 2018cni and 2020kyg}
\centering
\smallskip
\begin{tabular}{l  c c c c c}
\hline
SN 2018cni                          & B band           & V band           & g band            & r band              & i band  \\
\hline
JD of maximum light (2458000+)      & 289.99$\pm$0.5   & 289.99$\pm$0.5   & 289.99$\pm$0.5    & 290.00$\pm$0.5      &290.01$\pm$0.5     \\
Magnitude at maximum (mag)          & 18.36$\pm$0.07   & 18.03$\pm$0.04   & 18.13$\pm$0.02    & 17.98$\pm$0.03      & 18.09$\pm$0.08    \\
Absolute magnitude at maximum (mag) & $-$17.59$\pm$0.21  & $-$17.81$\pm$0.21  & $-$17.78$\pm$0.20   & $-$17.82$\pm$0.20    & $-$17.65$\pm$0.22    \\

\hline
$\Delta$m$_{15}$( mag)                  & 1.99$\pm$0.14    & 0.79$\pm$0.07     & 1.61$\pm$0.09     & 0.48$\pm$0.06    & 0.18$\pm$0.08   \\ 
\hline\hline
SN 2020kyg                          &          &            &          &              &  \\
\hline
JD of maximum light (2458000+)      & 998.94$\pm$2.0  & 996.88$\pm$0.5  & 998.95$\pm$2    & 1000.9$\pm$0.5      &1000.9$\pm$0.5     \\
Magnitude at maximum (mag)          & 18.36$\pm$0.06  & 18.28$\pm$0.14  & 18.37$\pm$0.02  & 18.31$\pm$0.03     & 18.38$\pm$0.04    \\
Absolute magnitude at maximum (mag) & $-$14.45$\pm$0.17 & $-$14.52$\pm$0.21  & $-$14.44$\pm$0.16 & $-$14.49$\pm$0.16     & $-$14.41$\pm$0.17    \\

\hline
$\Delta$m$_{15}$( mag) & 2.54$\pm$0.13    & 0.98$\pm$0.19     & 1.77$\pm$0.08      & 0.84$\pm$0.06    & 0.61$\pm$0.06   \\                       
\hline

\end{tabular}
\newline
\label{tab:decay_rate_18cni_20kyg}      
\end{table*}

\section{Distance and extinction}
\label{sec:Distance and extinction}

The host galaxy of SN 2018cni GALEXASC J150122.78-101044.8, is located at a redshift of 0.032 \citep{2018ATel11728....1Z}. The extinction due to the Milky Way in the direction of SN 2018cni is $E(B-V)$ = 0.09 mag  \citep{2011ApJ...737..103S} which yields A$_{V}$ = 0.278 mag using Cardelli extinction law \citep{1989ApJ...345..245C}. The spectra of SN 2018cni do not exhibit significant Na ID absorption at the redshift of the host galaxy, implying negligible host extinction. SN 2020kyg is located on the outskirts of the host galaxy NGC 5012, hence noticeable extinction from the host is not expected. The extinction due to the Milky Way is $E(B-V)$ = 0.012 mag \citep{2011ApJ...737..103S} which results in A$_{V}$ = 0.038 mag assuming R$_{v}$ = 3.1. We used the redshift dependent luminosity distance for GALEXASC J150122.78-101044.8 and NGC 5012 as 130.56$\pm$12.15 Mpc and 35.83$\pm$2.66 Mpc, respectively (assuming $H_0$ = 73 km s$^{-1}$ Mpc$^{-1}$, $\Omega_m$ = 0.27,  $\Omega_v$ = 0.73). 

\section{Light curve properties}
\label{sec:Light curve properties}

\subsection{Light curve and color curve}
\label{Light curve and colour curve}

\begin{figure}
	\begin{center}
		\includegraphics[width=\columnwidth]{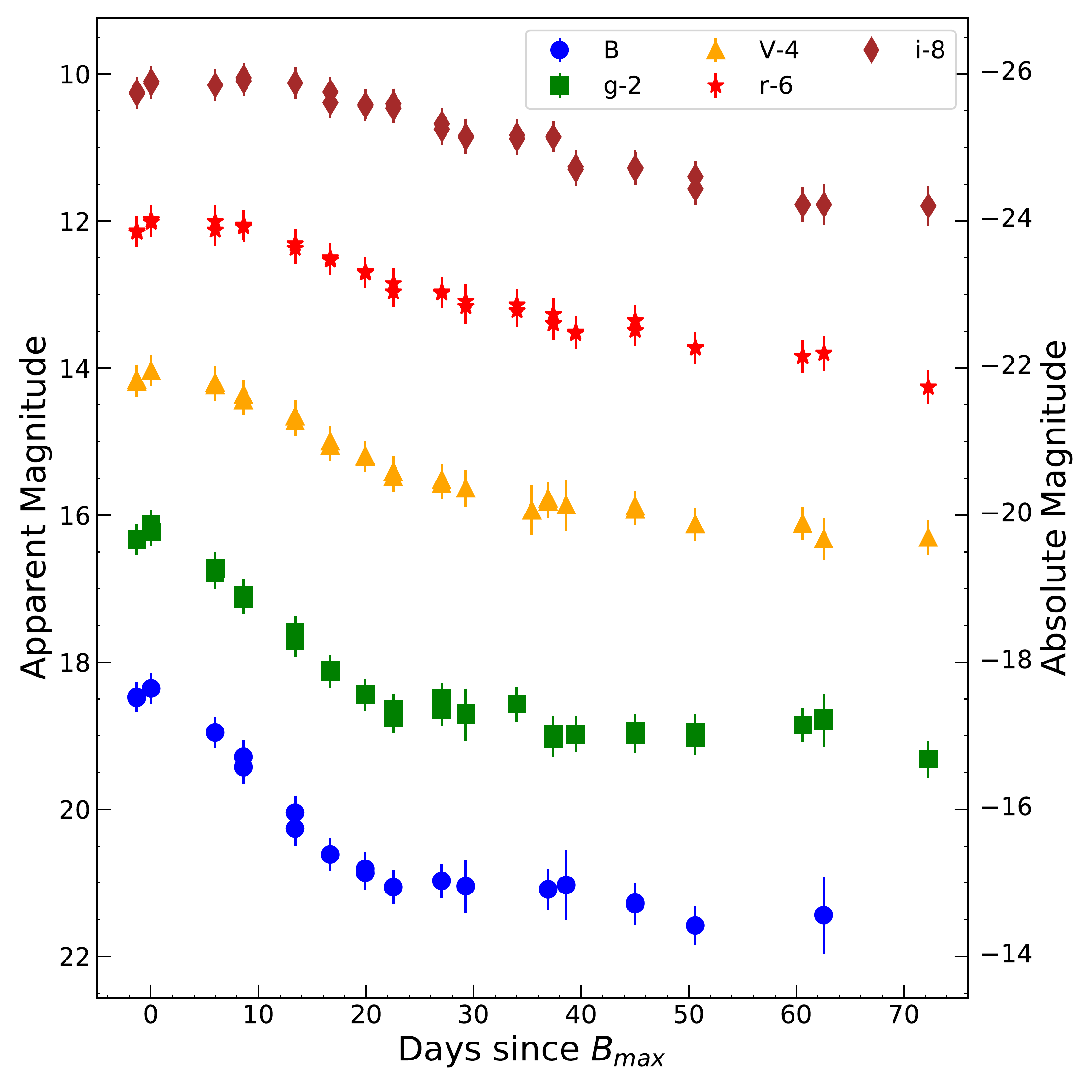}
	\end{center}
	\caption{This figure shows the light curve evolution of SN 2018cni in $BgVri$ bands. In the right Y-axis, absolute magnitudes in the corresponding $BgVri$ bands are presented.}
	\label{fig:SN 2018cni_light_curve}
\end{figure}

\begin{figure}
	\begin{center}
		\includegraphics[width=\columnwidth]{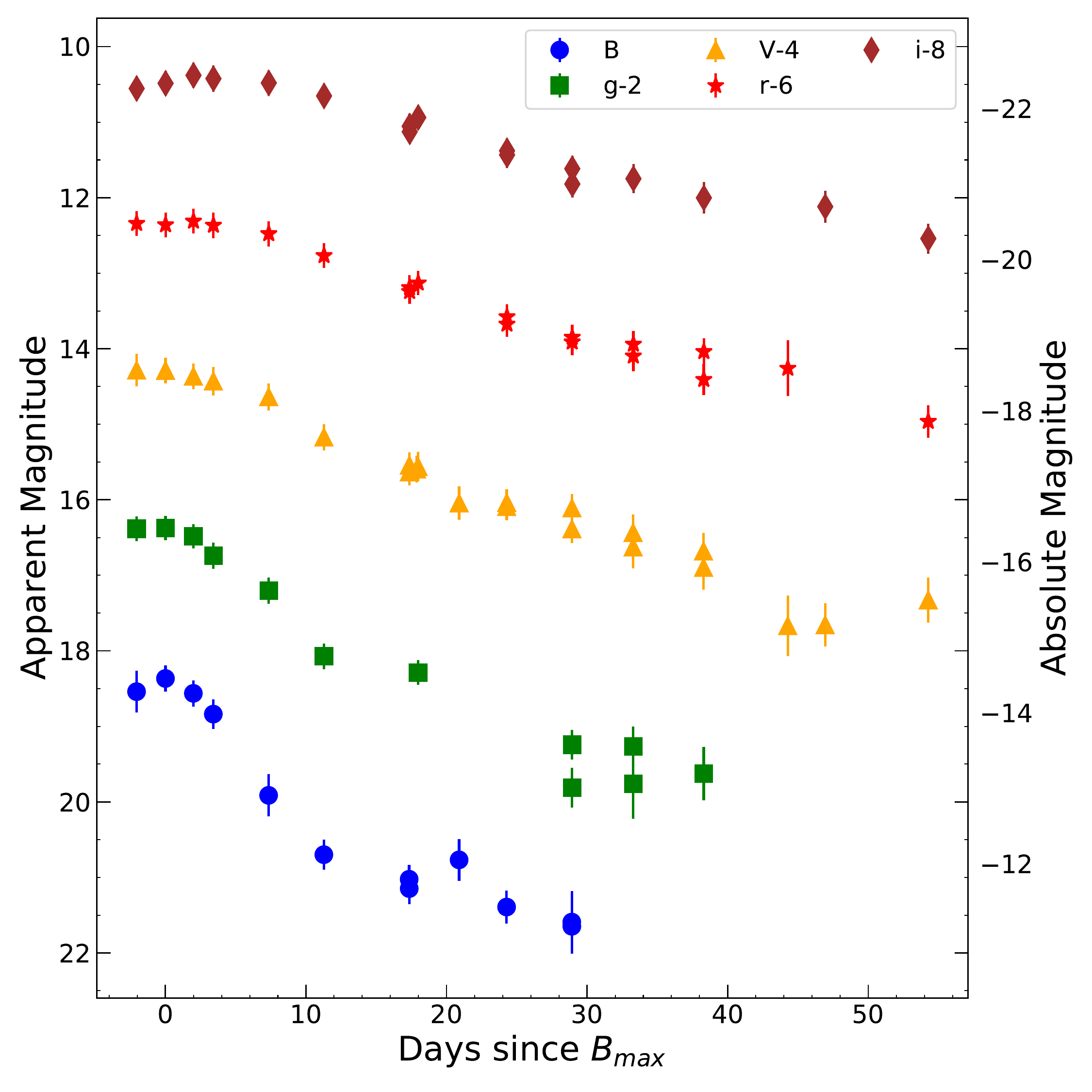}
	\end{center}
	\caption{Light curve evolution of SN 2020kyg in $BgVri$ bands. Absolute magnitudes in $BgVri$ bands for SN 2020kyg are shown on the right Y-axis.}
	\label{fig:SN 2020kyg_light_curve}
\end{figure}

\begin{figure}
	\begin{center}
		\includegraphics[width=\columnwidth]{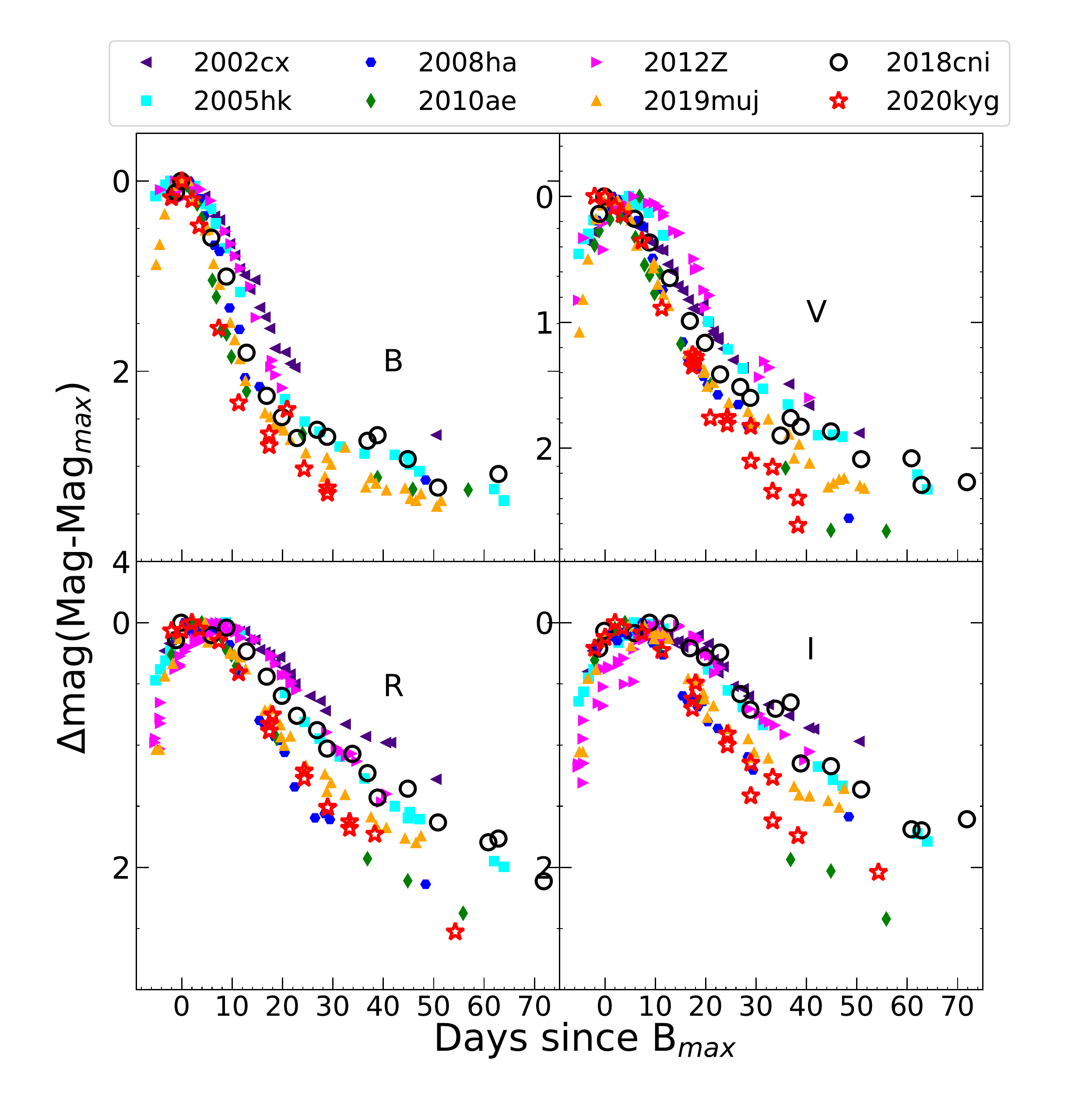}
	\end{center}
	\caption{Comparison of the normalized light curves of SNe 2018cni and 2020kyg with a few other well studied type Iax SNe.}
	\label{fig:com_plot_2018cni_2020kyg}
\end{figure}

Figures \ref{fig:SN 2018cni_light_curve} and \ref{fig:SN 2020kyg_light_curve} present the light curve evolution of SNe 2018cni and 2020kyg, respectively. Observations began before both the SNe attained peak brightness. The time and magnitude at maximum along with decline rates for both the SNe are estimated using a low order spline fit to the light curves. The values are listed in Table \ref{tab:decay_rate_18cni_20kyg}. Light curve decline rates, $\Delta$$m$$_{15}$($B$), measured for SNe 2018cni and 2020kyg in $B$-band are 1.99$\pm$0.14 mag and 2.54$\pm$0.13 mag, respectively. The $I$-band light curves of SNe 2018cni and 2020kyg do not show a secondary maximum, usually seen in type Ia SNe. Figure \ref{fig:com_plot_2018cni_2020kyg} presents the light curve comparison of SNe 2018cni and 2020kyg in  $BVRI$ bands with a few well-studied type Iax SNe 2002cx \citep{2003PASP..115..453L}, 2005hk \citep{2008ApJ...680..580S}, 2008ha \citep{2009AJ....138..376F}, 2010ae \citep{stritzinger2014}, 2012Z \citep{2015ApJ...806..191Y} and 2019muj \citep{2021MNRAS.501.1078B}, covering the luminosity range for this class. The $r$ and $i$-band magnitudes of SNe 2010ae, 2018cni, and 2020kyg are converted to $R$ and $I$-band using transformation equations given by \cite{2006A&A...460..339J}. The magnitudes of each SN in the respective bands are normalized by their peak magnitudes. In $B$ and $V$ bands, SN 2018cni declines faster than SNe 2002cx, 2005hk, and 2012Z, while SN 2020kyg is the fastest declining SN amongst all the SNe used for comparison as shown in Figure \ref{fig:com_plot_2018cni_2020kyg}. In $R$ and $I$-bands, the light curve evolution of SN 2018cni is similar to SNe 2005hk and 2012Z while SN 2020kyg resembles SN 2008ha. 

\begin{figure}
	\begin{center}
		\includegraphics[width=\columnwidth]{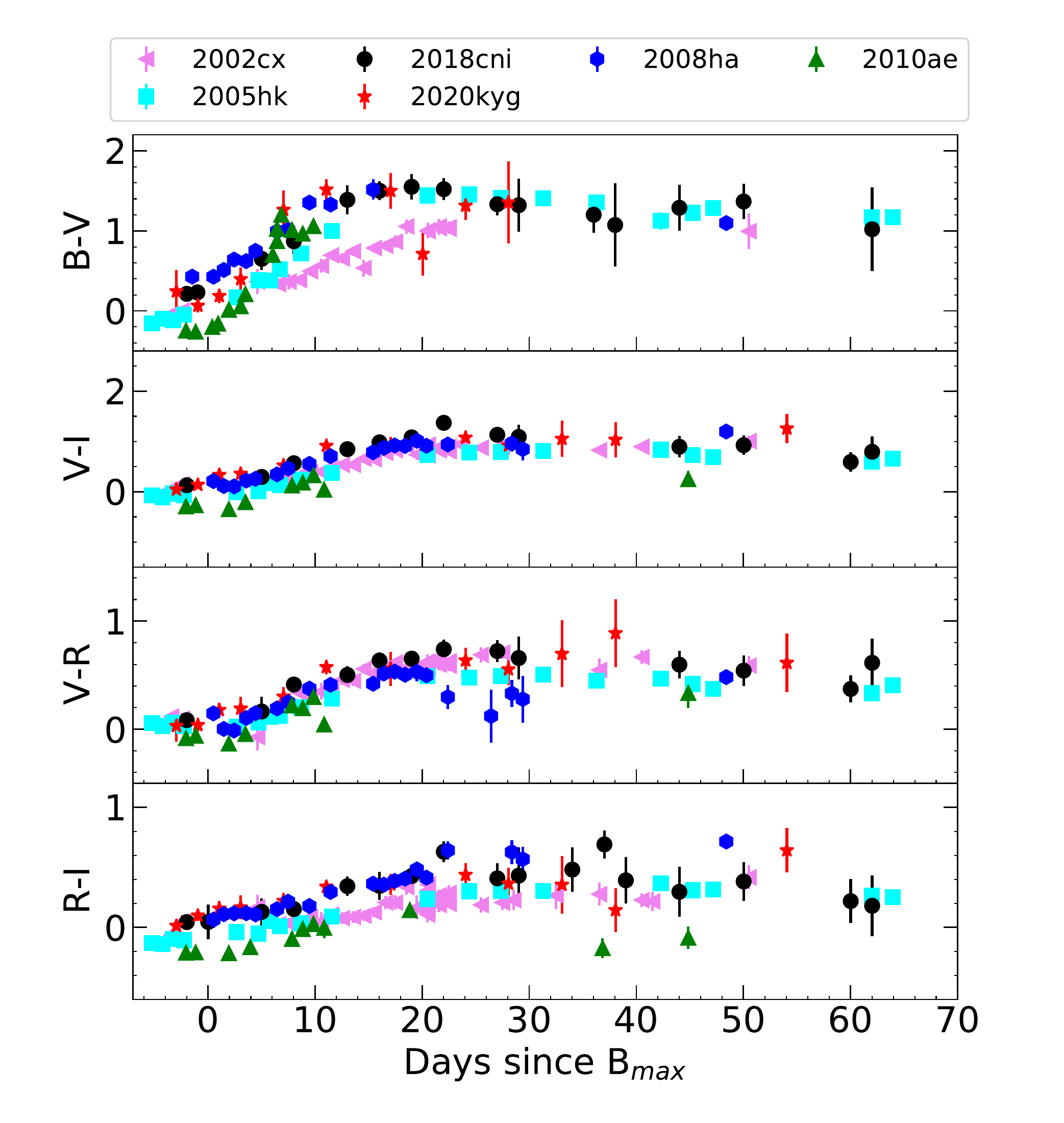}
	\end{center}
	\caption{Color curves of SNe 2018cni and 2020kyg and their comparison with color evolution of other type Iax SNe.}
    \label{fig:SN_2018cni_2020kyg_color_curve}
\end{figure}

Figure \ref{fig:SN_2018cni_2020kyg_color_curve} presents the reddening corrected color evolution of SNe 2018cni and 2020kyg and their comparison with other well studied type Iax SNe. The ($B$-$V$), ($V$-$I$), ($V$-$R$), and ($R$-$I$) color evolution of SNe 2018cni and 2020kyg  are found to be similar to the SNe used for comparison. This indicates that bright and faint members behave in a similar fashion in terms of color evolution. 

\subsection{Bolometric light curve and analytical modelling}
\label{sec:Bolometric light curve and analytical modelling}

\begin{figure}
	\begin{center}
		\includegraphics[scale=0.5, clip, trim={0.25cm 0.7cm 1.5cm 1.8cm}]{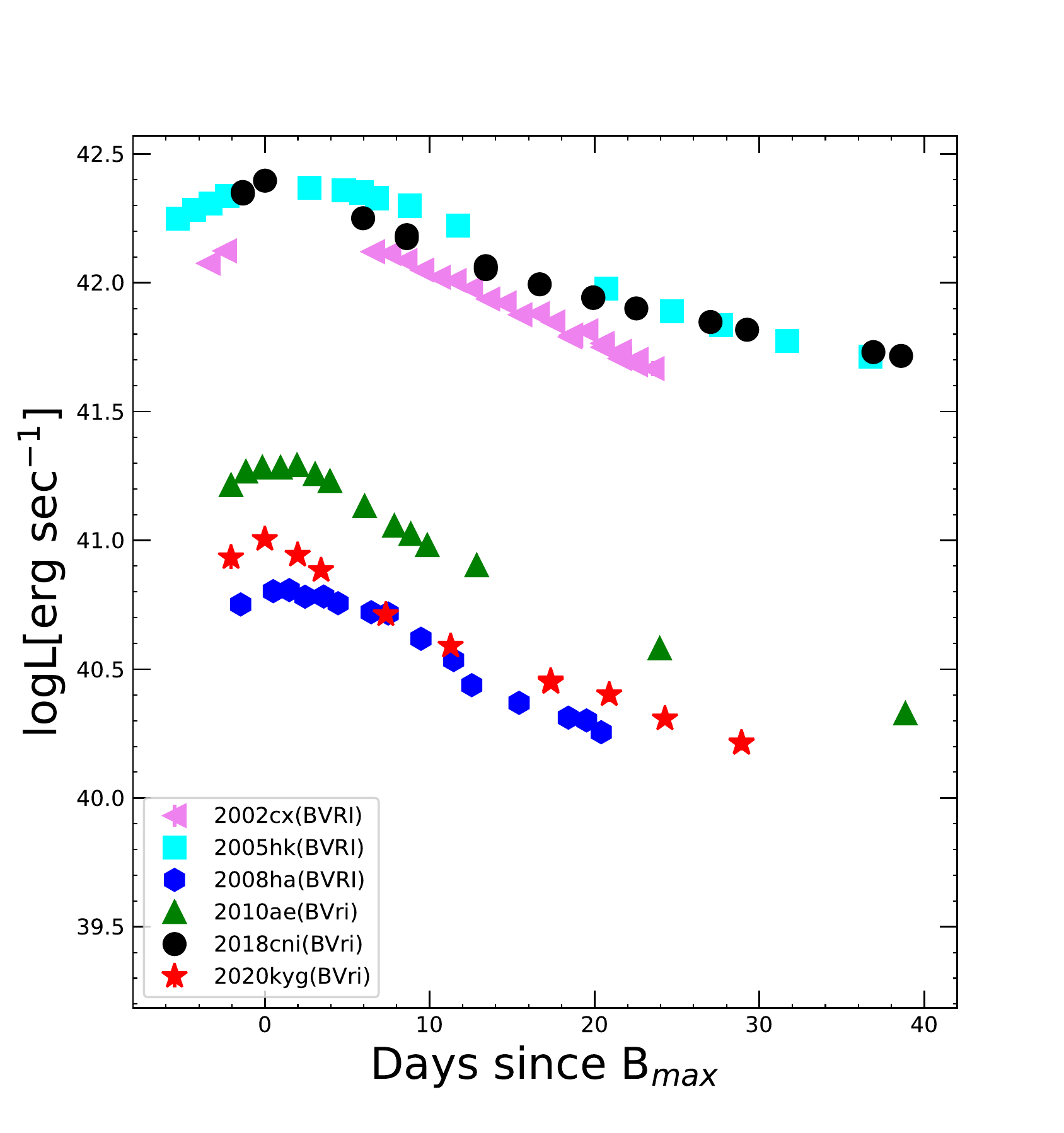}
	\end{center}
	\caption{ This figure presents the pseudo-bolometric light curves of SNe 2018cni and 2020kyg. The pseudo-bolometric light curves of SNe 2002cx, 2005hk, 2008ha, and 2010ae are included for comparison. The figure shows that SN 2018cni belongs to the bright type of Iax SNe whereas SN 2020kyg conforms to the fainter class of type Iax SNe.}
	\label{fig:SN 2018cni_2020kyg_bolometric_light_curve}
\end{figure}

\begin{figure}
	\begin{center}
		\includegraphics[scale=0.4, clip, trim={0.25cm 0.8cm 1.5cm 2.8cm}]{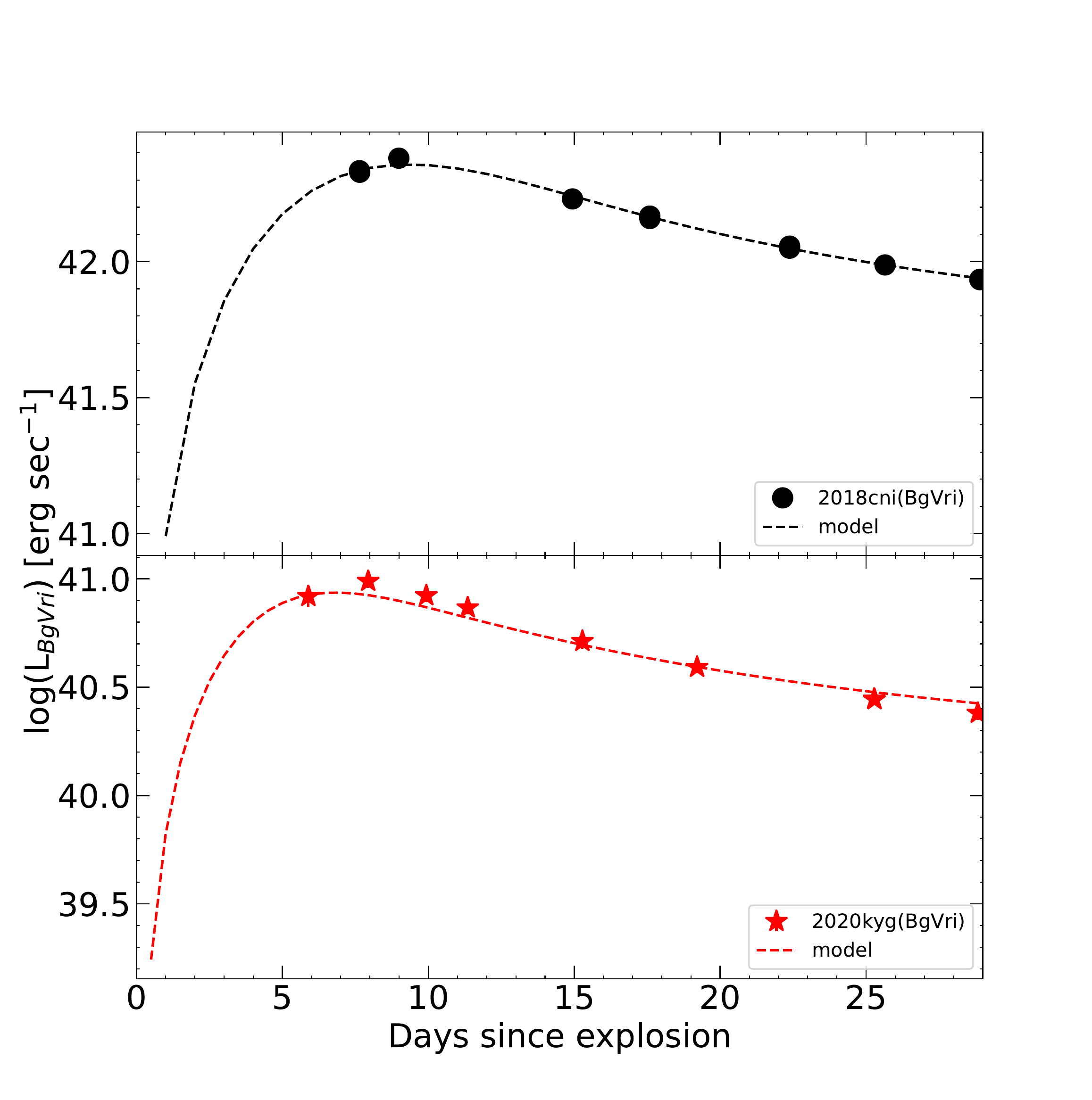}
	\end{center}
	\caption{Pseudo-bolometric light curves of SNe 2018cni and 2020kyg obtained through SuperBol along with their respective Arnett-Valenti model (dashed line) fits. }
	\label{fig:2018cni_2020kyg_com_bol_model_light_curve}
\end{figure}

The pseudo-bolometric light curves of SNe 2018cni and 2020kyg are constructed with \texttt{SuperBol} code \citep{2018RNAAS...2..230N} in $BVri$ bands using the luminosity distance discussed in Section \ref{sec:Distance and extinction}. The pseudo-bolometric light curves of SNe 2002cx, 2005hk, 2008ha, and 2010ae are constructed in a similar manner and are presented in Figure \ref{fig:SN 2018cni_2020kyg_bolometric_light_curve}. We find that the bolometric luminosity of SN 2018cni is similar to bright type Iax SN 2005hk whereas SN 2020kyg lies between SNe 2008ha and 2010ae and belongs to the fainter class of type Iax SNe. We have also used \texttt{SuperBol} code to calculate the $BgVri$ luminosities of SNe 2018cni and 2020kyg which are shown in Figure \ref{fig:2018cni_2020kyg_com_bol_model_light_curve}. 

Explosion parameters of both the SNe are estimated using Arnett's model \citep{1982ApJ...253..785A} along with the formulation given in \cite{2008MNRAS.383.1485V}.  Homologous expansion, spherically symmetric ejecta, no mixing of $^{56}$Ni and optically thick ejecta are basic assumptions used in the model. Mass of $^{56}$Ni and the timescale of the light curve ($\tau$) are free parameters for the fitting. Optical opacity is fixed to 0.07 cm$^{2}$g$^{-1}$ and photospheric velocities used are 7000 km s$^{-1}$ and 4500 km s$^{-1}$ (Section \ref{Spectral features and comparison}) for SNe 2018cni and 2020kyg, respectively. 

\begin{figure}
	\begin{center}
		\includegraphics[scale=0.42, clip, trim={0.85cm 0.6cm 1.5cm 2.4cm}]{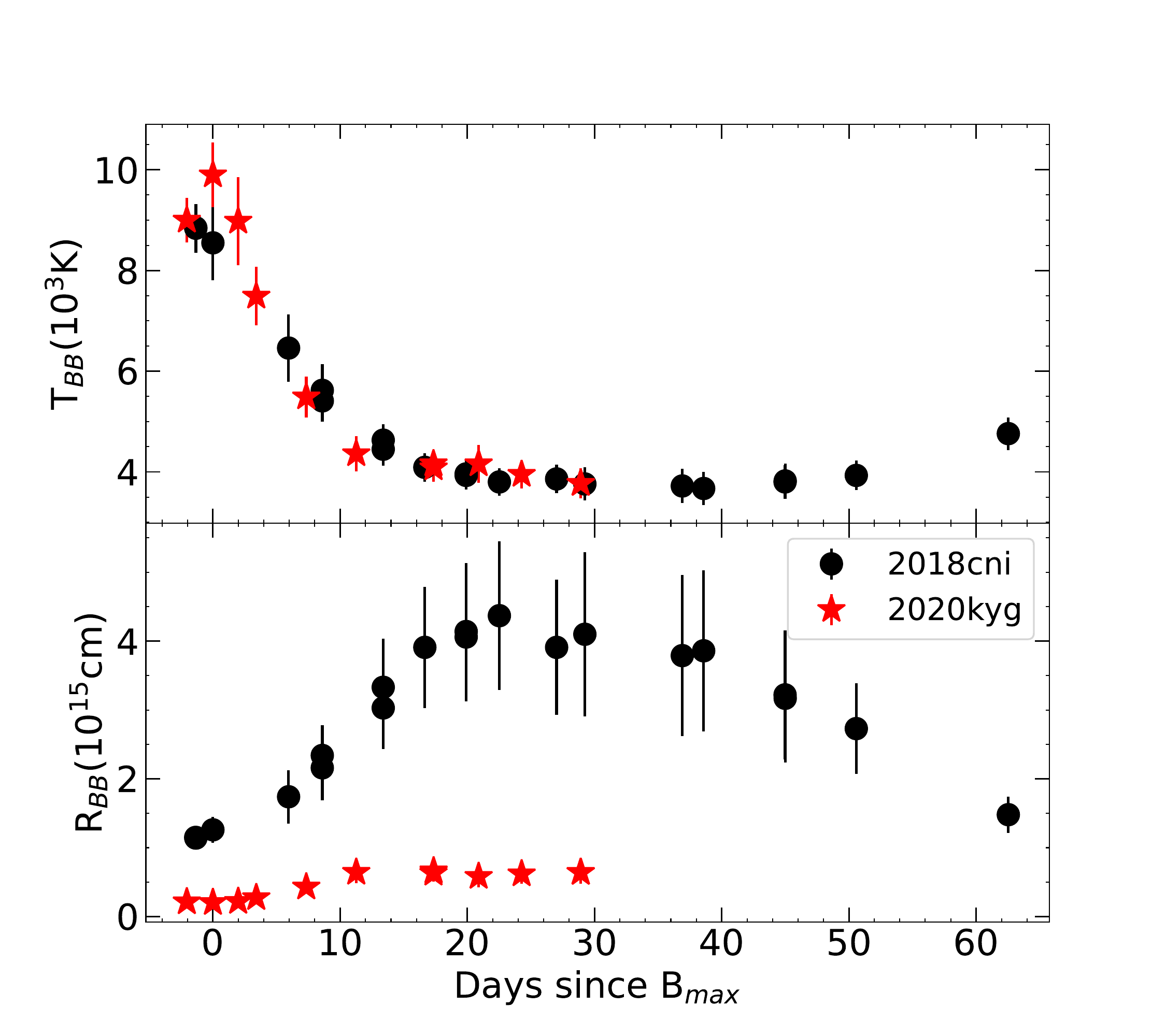}
	\end{center}
	\caption{Figure depicts the evolution of blackbody temperature and radius obtained through \texttt{SuperBol}. The evolution of the blackbody temperature is same for both SNe, whereas the blackbody radius is higher for SN 2018cni than SN 2020kyg.}
	\label{fig:Com_supebol_2018cni_2020kyg}
\end{figure}

Analytical modelling of the pseudo bolometric light curve of SN 2018cni gives mass of $^{56}$Ni $\sim$ 0.07$\pm$0.01M$_{\odot}$ and the ejecta mass of 0.48 M$_{\odot}$. For SN 2020kyg, these estimates are mass of $^{56}$Ni $\sim$ 0.002$\pm$0.001 M$_{\odot}$ and ejecta mass of 0.14 M$_{\odot}$. For SN 2020kyg, our estimated values are lower than that presented in \cite{2022MNRAS.511.2708S}, where {\it ugrizyJHK} fluxes were used for estimating the explosion parameters. The evolution of the blackbody temperature and radius for SNe 2018cni and 2020kyg are shown in Figure \ref{fig:Com_supebol_2018cni_2020kyg}. Despite the large difference in luminosity for the two SNe, their blackbody temperature evolution is similar whereas the blackbody radius is larger for the bright SN 2018cni. This suggests that diversity in luminosity is directly related to the radius of the photosphere which in turn is related to expansion velocity.   

Because of the intrinsic heterogeneity of SNe Iax,  the contribution of UV and NIR flux to the bolometric flux is not well constrained. However, depending on the availability of the data in UV and NIR bands, the fractional contribution of UV and NIR flux to the bolometric flux has been reported for a number of SNe Iax.  
For SN 2005hk, \cite{2007PASP..119..360P} estimated 20\% contribution of UV flux to the UVOIR light curve at the early time of evolution. \cite{2015ApJ...806..191Y} showed that the NIR band contributes $\sim$ 20\% to Opt+NIR  around maximum for SN 2012Z. A total UV+IR contribution of $\sim$ 35\% has been reported for SN 2014ck \citep{2016MNRAS.459.1018T}. \cite{2020ApJ...892L..24S} and \cite{2022ApJ...925..217D} quoted the ratio of peak quasi bolometric luminosity to peak black-body bolometric luminosity as 0.69 (SN 2019gsc) and 0.62 (SN 2020sck),  respectively. For SN 2020kyg, \cite{2022MNRAS.511.2708S} estimated a 60\% contribution from the optical luminosity to the total bolometric luminosity around the maximum. If $\sim$ 35\% flux is added to the pseudo-bolometric flux of SNe 2018cni and 2020kyg, the mass of $^{56}$Ni increases to 0.09 M$_{\odot}$  and 0.003 M$_{\odot}$, respectively. In this study, we have used pseudo-bolometric fluxes and reported the lower limit for the explosion parameters.

\begin{figure}
	\begin{center}
	\includegraphics[scale=0.35, clip, trim={0.7cm 1.1cm 1.5cm 3.0cm}]{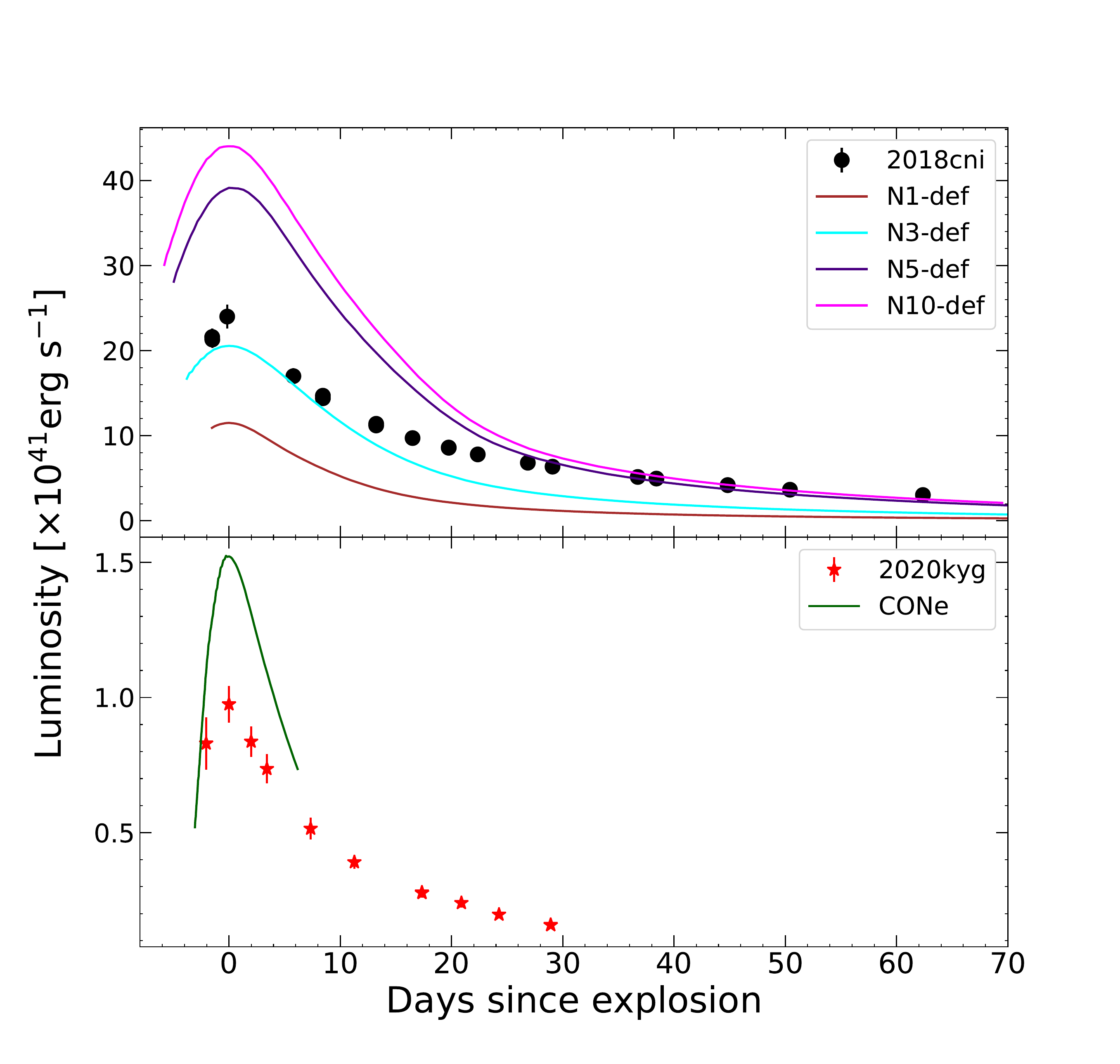}
	\end{center}
	\caption{Comparison of the bolometric light curves of different proposed models with the pseudo bolometric light curves of SNe 2018cni and 2020kyg.}
	\label{fig:deflag_2018cni_2020kyg}
\end{figure}

We compare the $BgVri$ bolometric light curves of SNe 2018cni and 2020kyg with model light curves of deflagration of CO white dwarf \citep{2014MNRAS.438.1762F} and hybrid CONe white dwarf \citep{2015MNRAS.450.3045K}, respectively in Figure \ref{fig:deflag_2018cni_2020kyg}. These model light curves are taken from the HESMA database. In the model light curves of CO white dwarfs, N1-def, N3-def, N5-def, and N10-def correspond to 1, 3, 5, and 10 ignition spots, respectively. The increasing number of ignition spots indicates a stronger explosion. We find that SN 2018cni is similar to the N3-def model and the low-luminosity SN 2020kyg is better explained by the deflagration of the CONe white dwarf. Since the models used for comparison include flux from UV to NIR and we have only $BgVri$ fluxes, we intend to only investigate the probable explosion scenario by finding the best match between the theoretical models and the observed luminosity of both the SNe.

\section{Prime spectral features}
\label{sec:Prime spectral features}

\subsection{Spectral features and comparison with other SNe}
\label{Spectral features and comparison}

\begin{figure}
	\begin{center}
		\includegraphics[scale=0.45, clip, trim={1.9cm 0.7cm 1.2cm 1.8cm}]{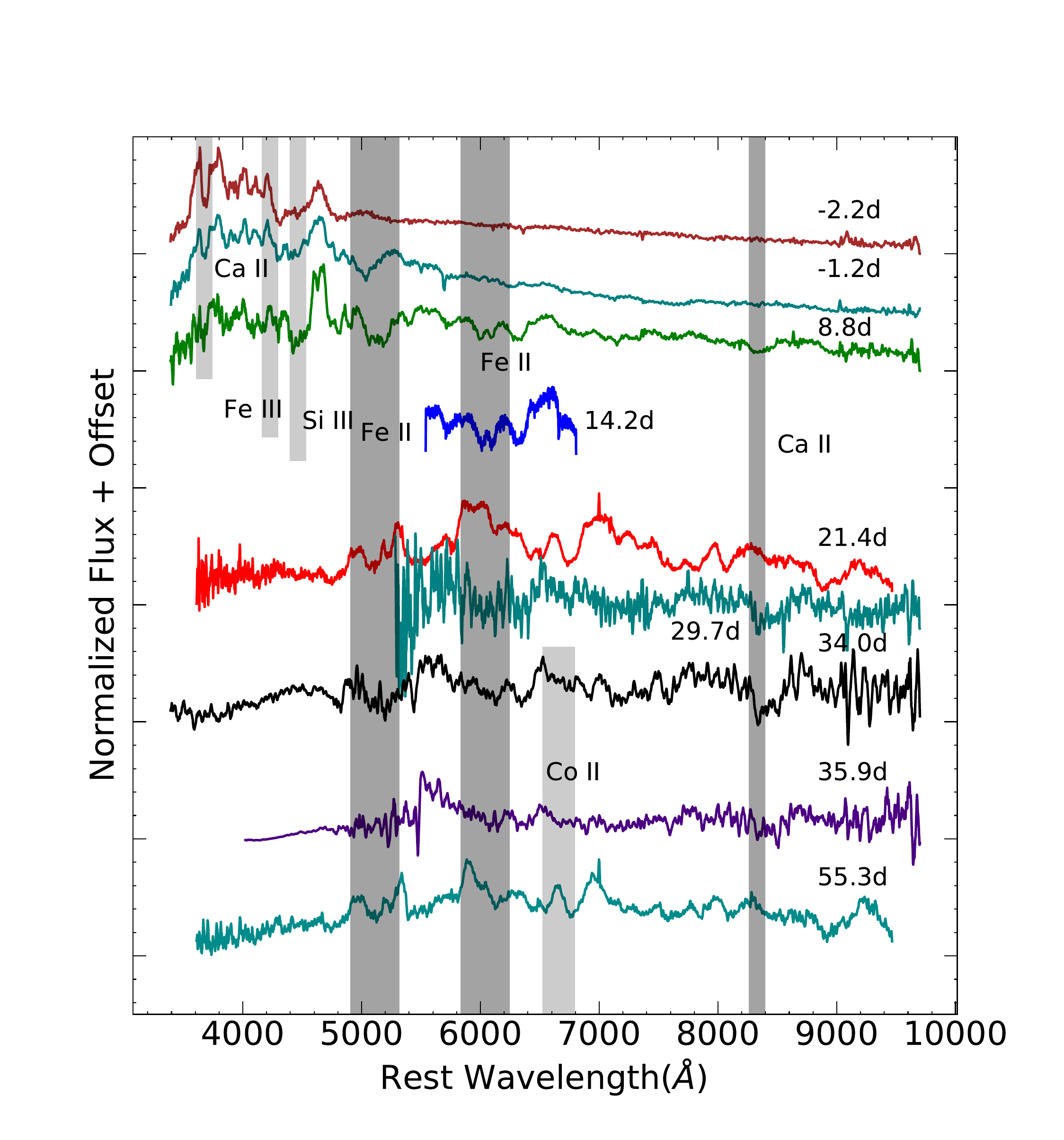}
	\end{center}
	\caption{Spectral evolution of SN 2018cni from $-$2 to 55 day since maximum.}
	\label{fig:SN 2018cni_spectral_sequence}
\end{figure}

\begin{figure}
	\begin{center}
		\includegraphics[scale=0.41, clip, trim={2.0cm 0.7cm 1.5cm 1.8cm}]{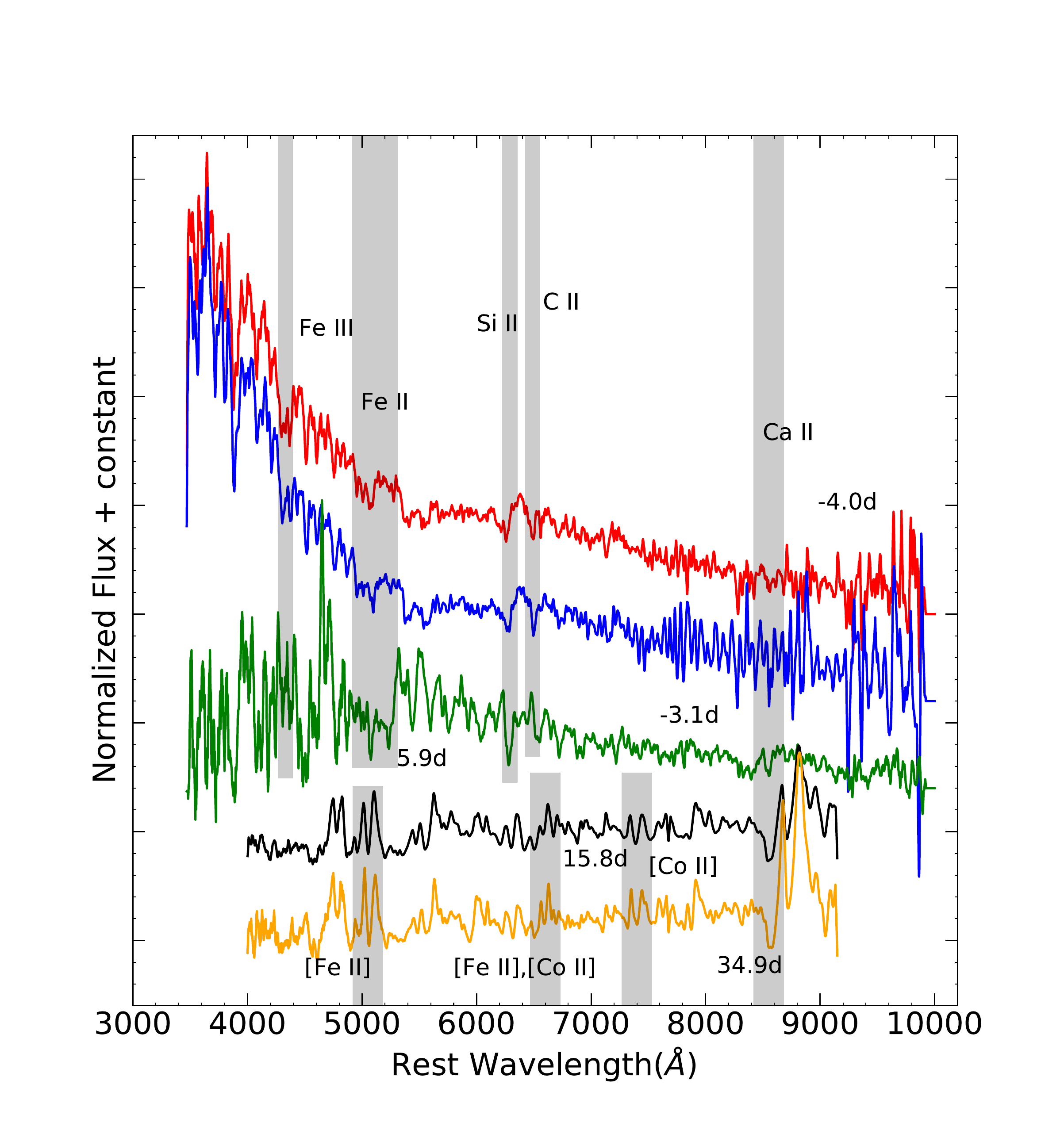}
	\end{center}
	\caption{ Spectral sequence of SN 2020kyg at $-$4.0, $-$3.1, 5.9, 15.8, and 34.9 day since maximum.}
	\label{fig:SN 2020kyg_spectral_sequence}
\end{figure}

\begin{figure*}
\centering
\begin{minipage}{.45\textwidth}
  \centering
  \includegraphics[width=\linewidth]{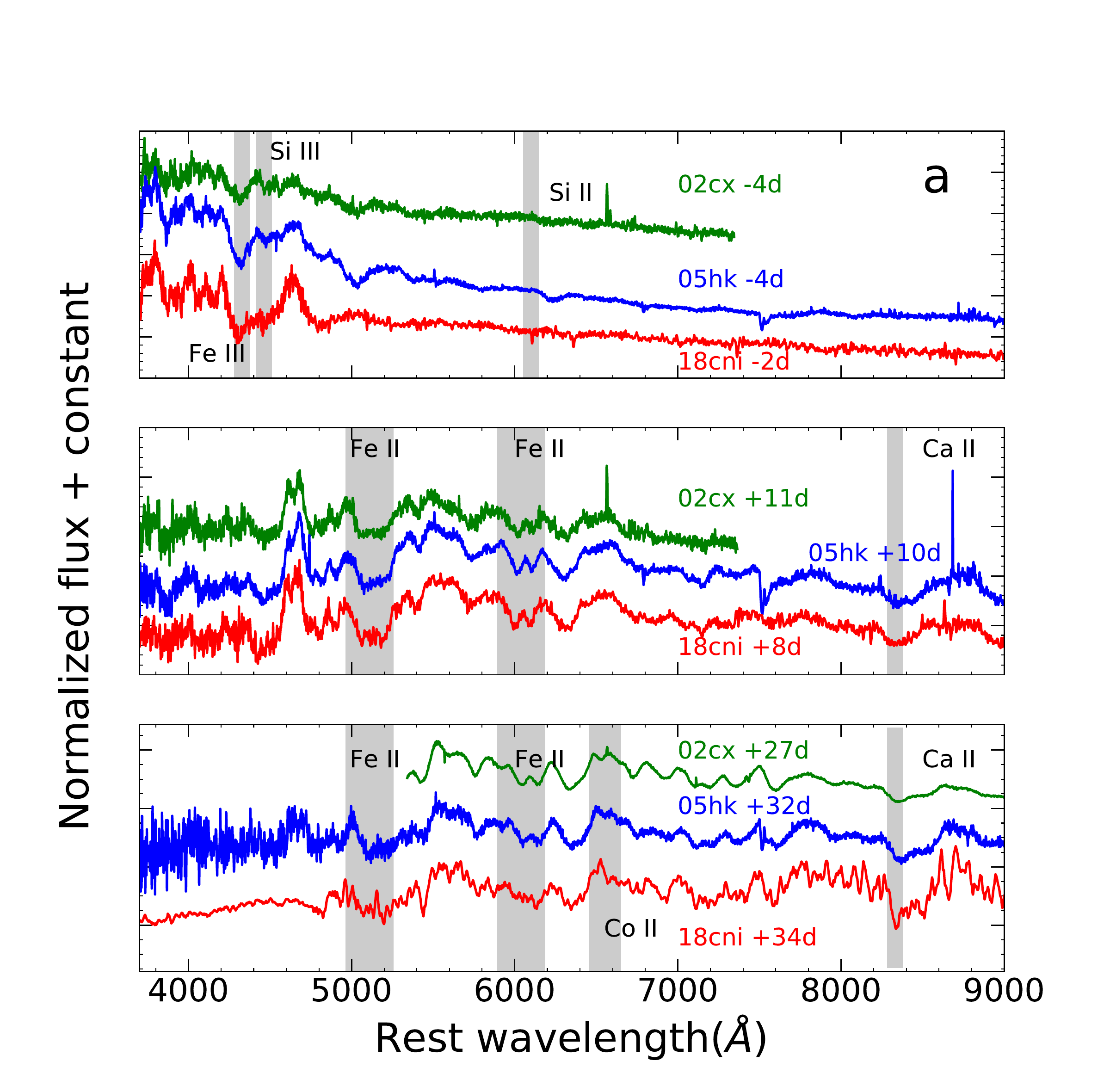}
  \label{fig:spectral_comp_2018cni}
\end{minipage}%
\begin{minipage}{.45\textwidth}
  \centering
  \includegraphics[width=\linewidth]{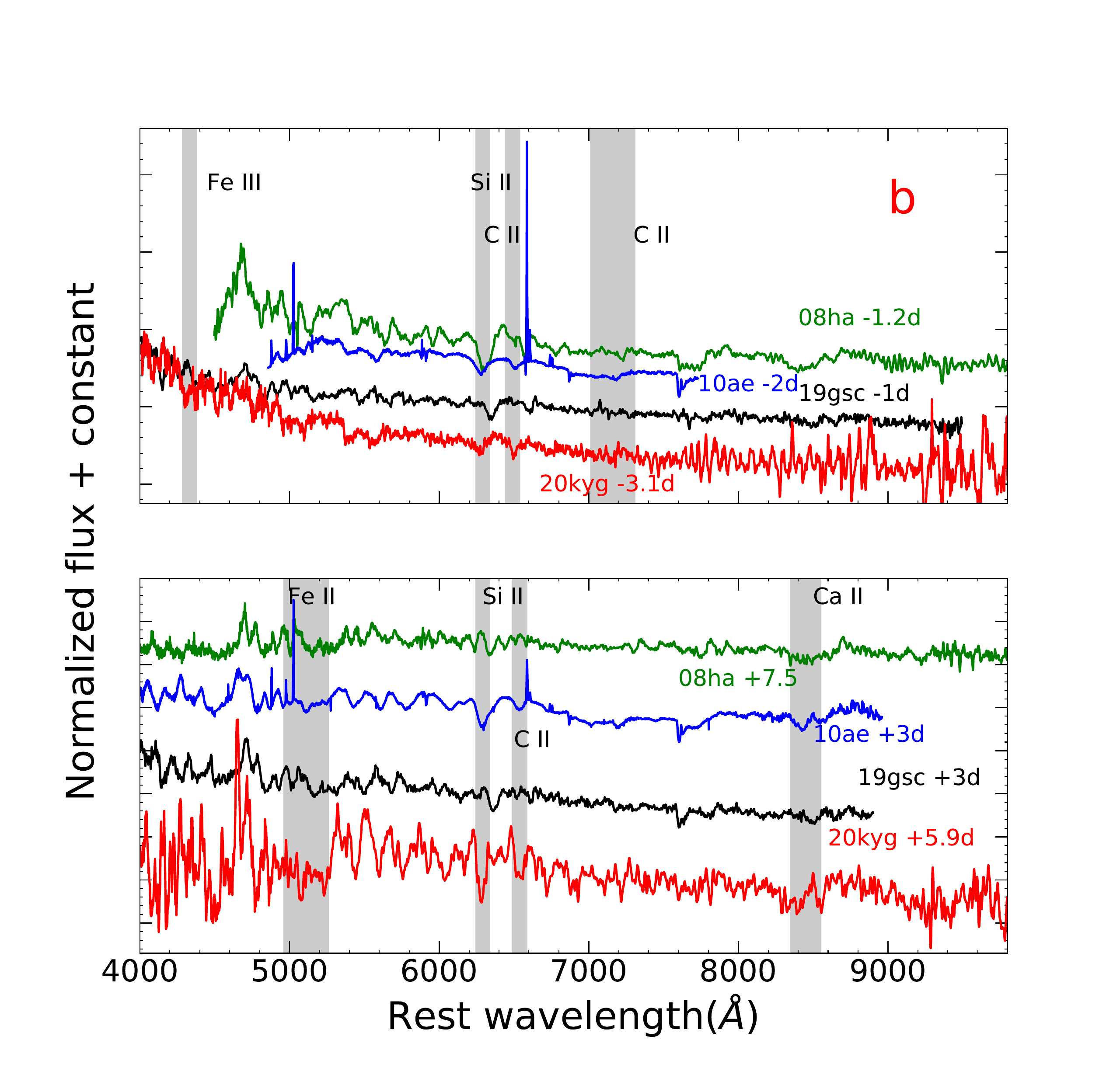}
  \label{fig:spectral_comp_2020kyg}
\end{minipage}
\caption{(a) This figure represents the comparison between spectra of SNe 2018cni, 2002cx, and 2005hk. (b) This figure shows the comparison between the spectra of SNe 2020kyg, 2008ha, 2010ae, and 2019gsc. The comparison features in both the figures are marked with shaded bars.}
\end{figure*}

Figures \ref{fig:SN 2018cni_spectral_sequence} and \ref{fig:SN 2020kyg_spectral_sequence} show the spectral sequence of SNe 2018cni and 2020kyg from $-$2 day to $+$55 day and $-$4.0 day to $+$35 day with respect to the maximum in $B$--band, respectively. The first spectrum of SN 2018cni at $-$2 day shows a blue continuum with \ion{Ca}{2} H\&K, \ion{Fe}{3}, and \ion{Si}{3} lines, while beyond 5200 \AA, the spectrum is featureless. In the post-maximum phase, the continuum in the blue region weakens. The post-maximum spectra of SN 2018cni exhibit \ion{Fe}{2} features near 5000 \AA, \ion{Co}{2} feature near 6500 \AA~ and \ion{Ca}{2} NIR feature. Spectra of SN 2018cni are compared with spectra of SNe 2002cx \citep{2003PASP..115..453L} and 2005hk \citep{2008ApJ...680..580S}, at comparable epochs in Figure \ref{fig:spectral_comp_2018cni}. In the pre-maximum phase (top panel), the  strength of \ion{Fe}{3} $\lambda$4420 line appears to be similar in all the three SNe while \ion{Si}{3} $\lambda$4553, 4568 lines in SN 2018cni appear stronger than the other two SNe. During the post-maximum epoch (middle and bottom panel), spectra of SN 2018cni are dominated by Iron Group Elements (IGE) and display a good one-to-one correspondence in the spectral features with SNe 2002cx and 2005hk. 

In the spectral evolution of SN 2020kyg, narrow P-Cygni profiles of Intermediate Mass Elements (IMEs) and IGEs are present. In Figure \ref{fig:spectral_comp_2020kyg}, we compare the spectra of SN 2020kyg with other faint type Iax SNe 2008ha \citep{2009AJ....138..376F}, 2010ae \citep{stritzinger2014} and 2019gsc \citep{2020ApJ...892L..24S}. In the pre-maximum phase, strong features due to \ion{C}{2} and \ion{Si}{2} can be seen in all the spectra. Presence of \ion{C}{2} is indicative of unburnt Carbon in the progenitor system. In the early post-maximum phase, \ion{C}{2} and \ion{Si}{2} lines become weak, and \ion{Fe}{2} around 5000 \AA~ can be seen with almost equal strength in spectra of SN 2020kyg and other SNe used for comparison. In the post maximum phase, \ion{Ca}{2} NIR feature starts developing in all the SNe. Overall similarity can be seen in the spectral signatures of SN 2020kyg with other SNe used for comparison. Spectra of SN 2020kyg at $+$16 day and $+$35 day are very similar, mainly dominated by permitted and forbidden lines due to \ion{Fe}{2}, \ion{Co}{2}, and \ion{Ca}{2}. The presence of forbidden lines in the $+$16 day spectrum indicates a considerably fast evolution of SN 2020kyg and its early transition into the nebular phase.

We measure the expansion velocity of SN 2020kyg by fitting a Gaussian function to the absorption minima of \ion{Si}{2} profile. The estimated velocities at three epochs ($-$4.0, $-$3.1, and $+$5.9 days) are 4330, 4290, and 3580 km s$^{-1}$, respectively. The measurement errors associated with the estimation of expansion velocities are $\sim$ 600 km s$^{-1}$. Our estimations are consistent with the values given in \citet{2022MNRAS.511.2708S}. As the identification of \ion{Si}{2} line in SN 2018cni is questionable, we did not attempt to measure its expansion velocity.  
\subsection{Spectral modelling}
\label{sec:Spectral modelling}

\begin{figure}
	\begin{center}
		\includegraphics[width=\columnwidth]{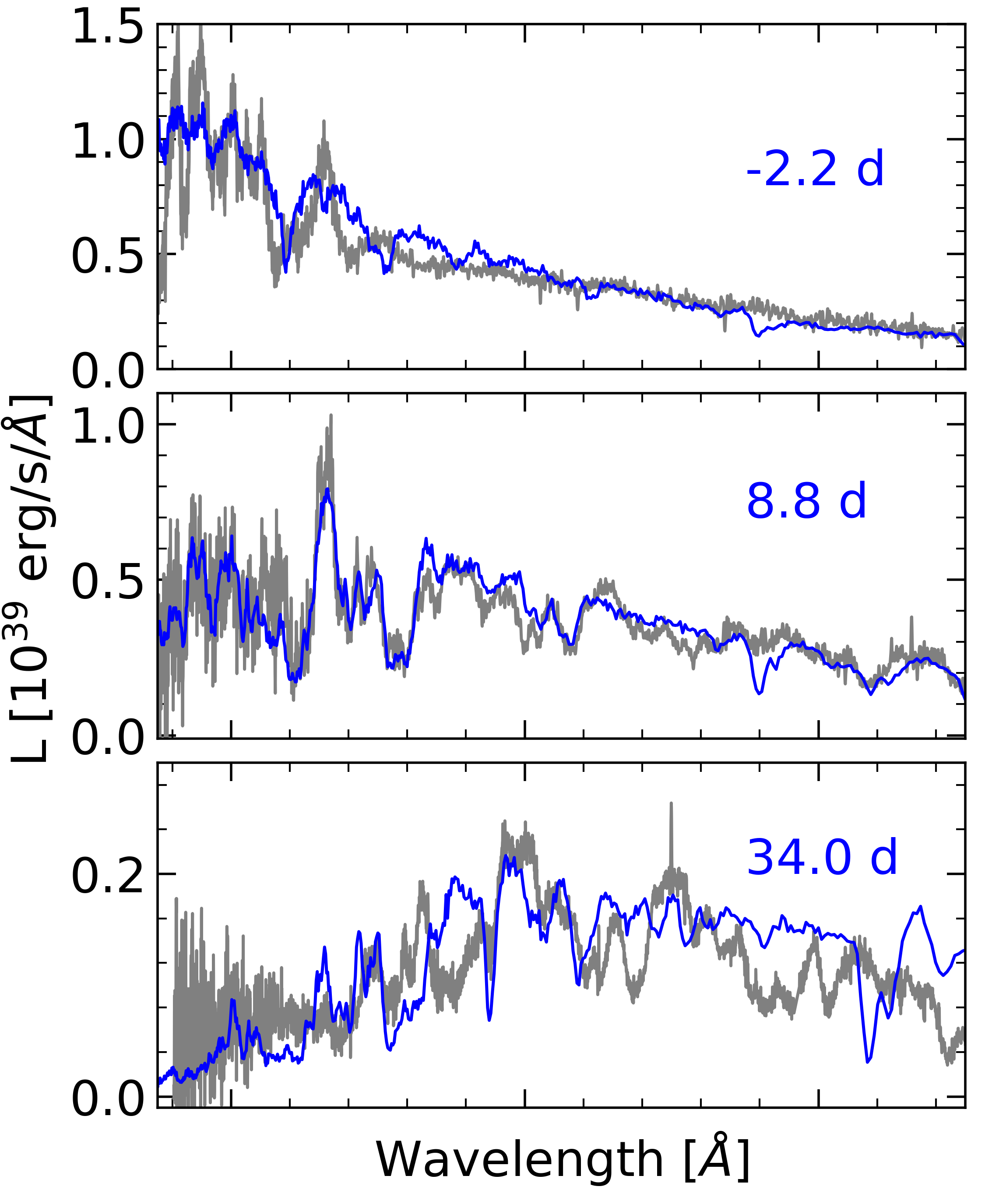}
	\end{center}
	\caption{This figure presents the spectral fitting of SN 2018cni using TARDIS. The grey and the blue lines show the SN spectra and the corresponding models, respectively.}
	\label{fig:18cni_plot_tardis}
\end{figure}
\begin{figure}
	\begin{center}
		\includegraphics[width=\columnwidth]{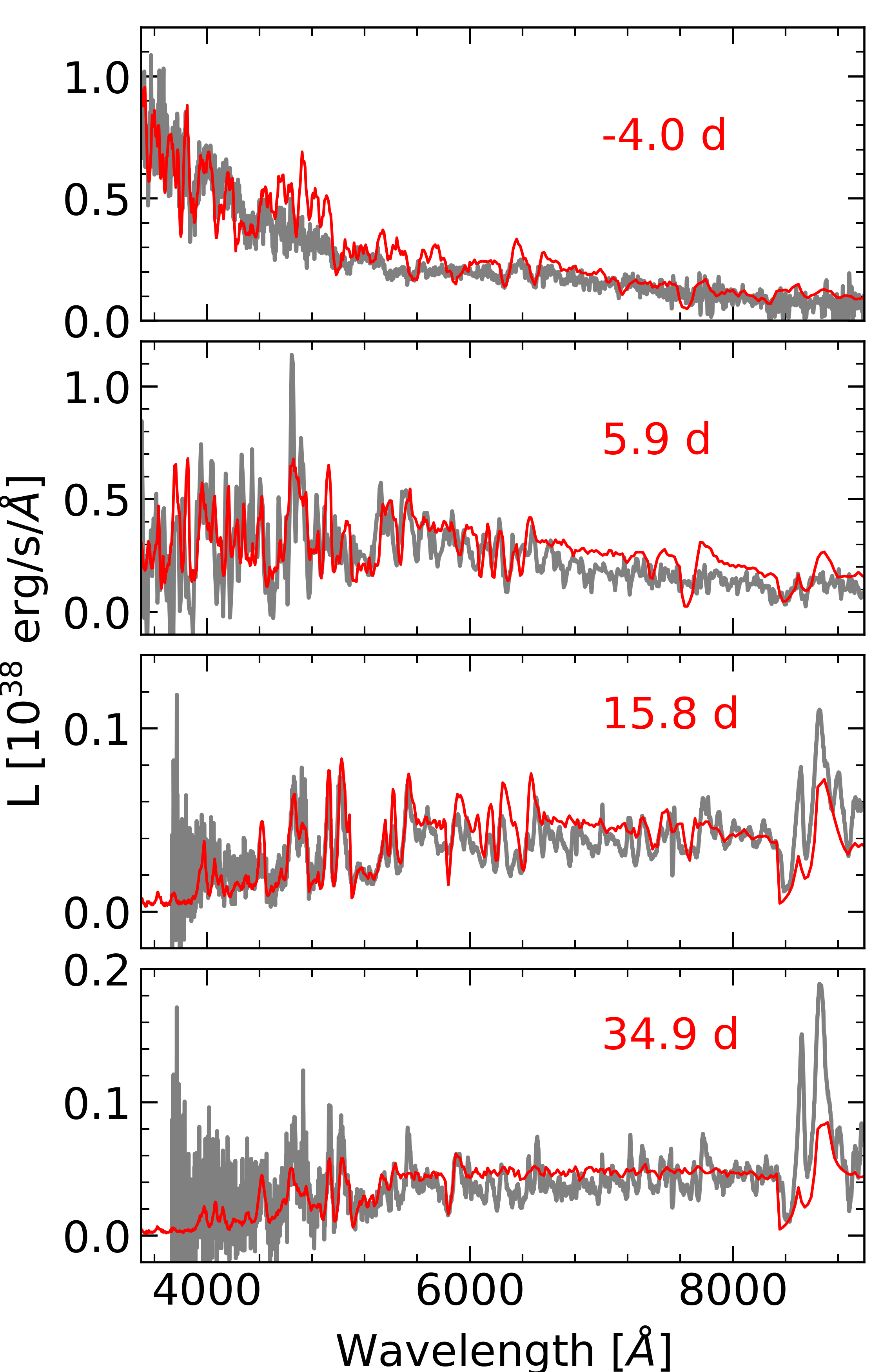}
	\end{center}
	\caption{Fitting of the spectra of SN 2020kyg using TARDIS is shown in this figure. Here, grey lines display the spectra of the SN and red lines present the respective models.}
	\label{fig:20kyg_plot_tardis}
\end{figure}

\begin{figure}
	\begin{center}
		\includegraphics[width=\columnwidth]{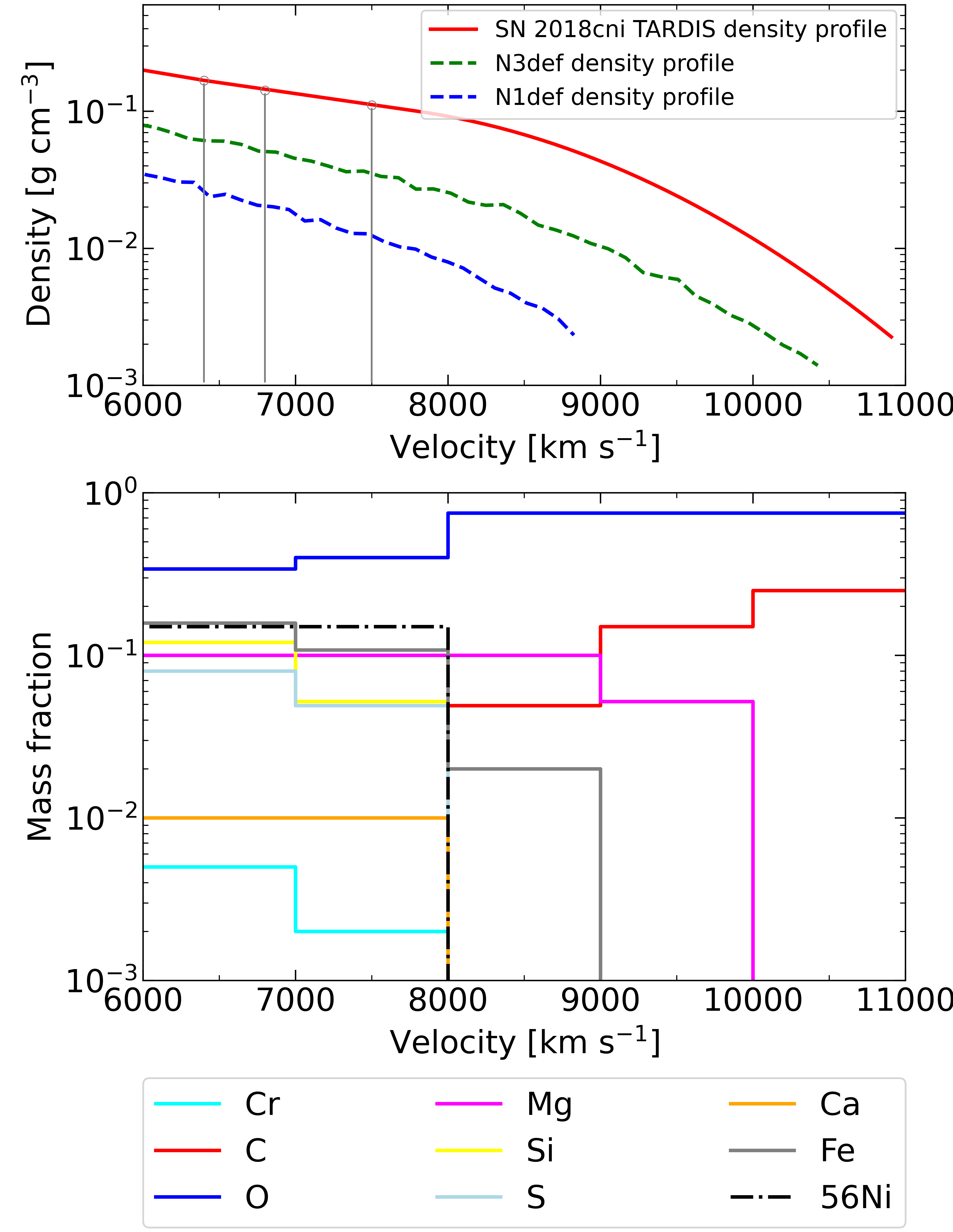}
	\end{center}
	\caption{\textit{Top panel:} The best-fit function for SN 2018cni (red) compared to the N1def and N3def pure deflagration density structure at $t_{exp} = t_0 = 100$ s. The grey lines indicate the boundaries of the fitting layers for the corresponding chemical abundance models. \textit{Bottom panel:} The best-fit chemical abundance structure from the TARDIS fitting process for the spectral sequence of SN 2018cni. The profile of the radioactive $^{56}$Ni shows the mass fractions at $t_{exp}=100$ s.}
\label{fig:abundance_density_2018cni} 
\end{figure}

\begin{figure}
	\begin{center}
 \includegraphics[width=\columnwidth]{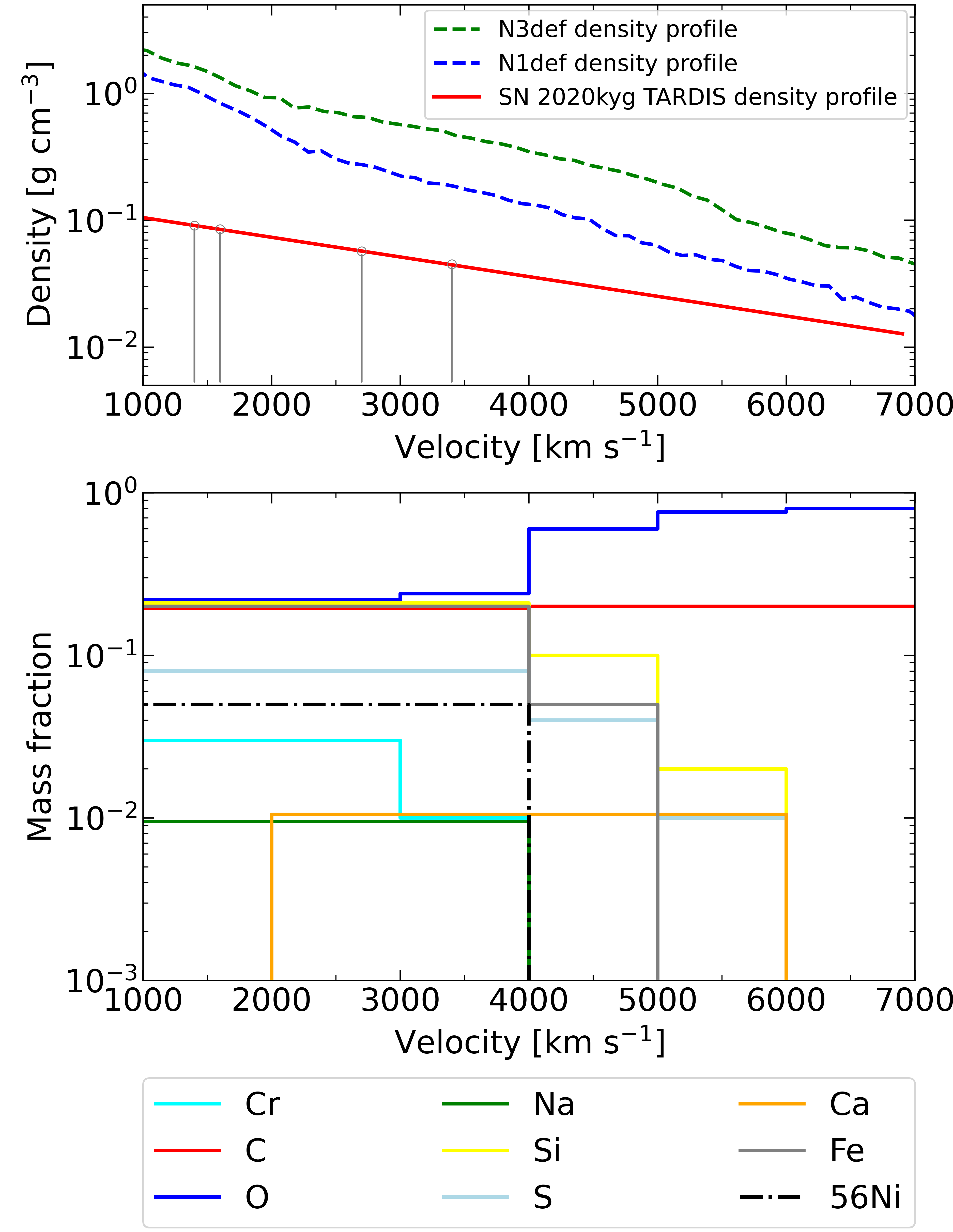}
	\end{center}
	\caption{Same as in Figure \ref{fig:abundance_density_2018cni} but for SN 2020kyg.}
	\label{fig:abundance_density_2020kyg}
\end{figure}

We perform spectral modelling for SN 2018cni at three epochs and SN 2020kyg at four epochs (Figures \ref{fig:18cni_plot_tardis} and \ref{fig:20kyg_plot_tardis}) using the one-dimensional radiative transfer code TARDIS \citep{2014MNRAS.440..387K,kerzendorf_wolfgang_2018_1292315}. TARDIS assumes a blackbody emitting photosphere with a custom designed, spherically symmetric ejecta. The one dimensional density and abundance structure were subject of fitting, following the abundance tomography method described in \cite{2018MNRAS.480.3609B}: as the homologously expanding SN ejecta becomes optically thinner, the velocity of the photosphere decreases and observations can probe deeper into the ejecta.

The assumed density function is designed to resemble those of the various pure deflagration models \citep{2014MNRAS.438.1762F}, similar to previous attempts of fitting SN Iax spectra \citep{2021MNRAS.501.1078B}. This density function includes a simple exponential inner and a more strongly decreasing outer part:
\begin{equation}
\resizebox{\columnwidth}{!}
{$
\rho (v,t_{{exp}}) = 
    \begin{cases}
    \rho_0 \cdot \left(\frac{t_{{exp}}}{t_0}\right)^{-3} \cdot \exp\left({-\frac{v}{v_0}}\right)
        & \mathrm{for}  ~v \leq v_{cut} \\
    \rho_0 \cdot \left(\frac{t_{{exp}}}{t_0}\right)^{-3} \cdot \exp\left({-\frac{v}{v_0}}\right) \cdot
    8^{-\frac{8(v - v_{cut})^2}{v^{2}_{cut}}}
        & \mathrm{for}~v > v_{cut}
    \end{cases}    
$
}
\end{equation}  

\noindent
where the innermost density $\rho_{0}$ (at reference time $t_{0} = 100$ s), the slope of the function $v_{0}$ and the start of the shallow cut-off past the velocity $v_{cut}$ are free parameters in the fitting process.

Abundances are constrained by fitting the prominent P-Cygni features at all spectral epochs of the actual SN, simultaneously. The model ejecta is split into 1000\,km\,s$^{-1}$ zones, where the mass fractions of the prominent chemical elements are fitting parameters. 

Due to the high number of mass fraction parameters, the whole parameter space cannot be fully explored manually. Thus, a different combination of physical and chemical properties, which would provide a similarly good match with the observed spectra, cannot be fully ruled out. However, considering that the main spectral features and the flux continuum of each epoch are well reproduced, we conclude that our results provide at least feasible solutions.

Other input parameters are the luminosity and the time since the explosion. The latter parameter controls the dilution of the density structure, and thus, affects the estimated temperature profile of the model ejecta. In the absence of good photometric coverage during the early phase, the explosion date is constrained from the spectral synthesis, as $t_{exp}$ has a critical impact on the pre-maximum epochs. The best-fit date is considered as the date of the explosion. In the case of SN 2020kyg, the inferred date of explosion is in agreement with that of \cite{2022MNRAS.511.2708S} (JD = 2458988.3) within the estimated $\pm1.0$ days uncertainty. The explosion epoch for SN 2018cni estimated in a similar manner is JD = 2458275.3. Luminosities are measured from the bolometric light curve, allowing variation within the uncertainty range of the bolometric estimations.

\begin{table}
\caption{Parameters of the best-fit models of SNe 2018cni and 2020kyg from our spectral synthesis method. $\Delta$t and v$_{max}$ represent t$_{exp}$ and v$_{phot}$ at maximum light.}
\smallskip
\begin{tabular}{c c c}
\hline \hline
&  SN 2018cni    &  SN 2020kyg   \\
\hline
$\Delta$t [days]     &    16     &   10.8    \\
v$_{max}$ [km s$^{-1}$] &  7500   & 3100    \\
\hline
Core density [g cm$^{-3}$]     & 1.40  & 0.15   \\
Density slope [km s$^{-1}$]    & 2700  & 2800  \\
Cut-off velocity [km s$^{-1}$]  &  7000  & - \\

\hline                                   
\end{tabular}
\newline
$^\dagger$ Assuming epoch of maximum light B$_{max}$= 2458289.99 and 2458998.94 JD for SNe 2018cni and 2020kyg, respectively
\label{tab:tardis_18cni_20kyg}      
\end{table}

The main parameters of the best-fit model structures for SNe 2018cni and 2020kyg can be found in Table \ref{tab:tardis_18cni_20kyg}. The two objects fit in the general trend of other SNe Iax studied by abundance tomography, which displays increasing ejecta mass and rise time with luminosity. The constrained density functions (Figures \ref{fig:abundance_density_2018cni} and \ref{fig:abundance_density_2020kyg}) exhibit a relatively good agreement with the predictions of the pure deflagration models with corresponding luminosities. 

\cite{2022MNRAS.511.2708S} fitted a different subset of the spectral time series of SN 2020kyg, following the same principles of the abundance tomography technique and the same radiative transfer code. The density functions show nearly the same profile with a slope of $v_0=2800$ km\,s$^{-1}$ (3000 km\,s$^{-1}$ in \citealt{2022MNRAS.511.2708S}) which is the key parameter in reproducing the observed P-Cygni features. The model of \cite{2022MNRAS.511.2708S} has also a slightly lower core density of $\rho_0 = 0.03$\,g\,cm$^{-3}$. As a significant difference, our model fitting constrained lower $v_{phot}$ velocities over the epochs (assuming a linear decrease with time): for the only epoch included in both samples ($t_{exp} = 6.5$ days), we derived $v_{phot} = 3500$ km\,s$^{-1}$ instead of 4200 km\,s$^{-1}$ which results in a hotter photosphere. The source of discrepancy is in the different characterization of the temperature profile, which also greatly affects the inferred chemical abundances. In the nearly uniform chemical structure of \cite{2022MNRAS.511.2708S}, carbon is the dominant element ($X(C) = 0.60$) with significant contributions of O and minor abundances of Si, Fe and $^{56}$Ni. Note that such an extreme abundance of C is in conflict with the plausible explosion scenarios.

In this study, we aimed to constrain the abundance structure by adopting the general characteristics of the pure deflagration models \citep{2014MNRAS.438.1762F, 2022A&A...658A.179L} as a starting point of our fitting process. The final results are plotted in Figures \ref{fig:abundance_density_2018cni} and \ref{fig:abundance_density_2020kyg}. The outer layers of the models are dominated by C and O, suggesting that not the entire ejecta was reached by the fusion flame. At lower velocities, O is still the most abundant element in the entire regime of our models, while carbon are excluded from the inner regions (the lower limit is 10000 km s$^{-1}$ for SN 2018cni and 4000 km s$^{-1}$ for SN 2020kyg), otherwise, the \ion{C}{2} lines would be extreme around the maximum light. These results strongly contradict the predictions of the pure deflagration scenario, but one should treat them with caution: O is also used as a puffer element to fill the not-constrained mass fractions, thus, X(O) is highly uncertain. 

In general, the chemical distribution of the inner structure (i.e. below the carbon-populated layers) resembles the predictions of the pure deflagration models \citep{2014MNRAS.438.1762F}, as the elements forming the dominant spectral features (O, Si, S, Fe, Ni) show nearly constant abundances. The total IGE mass fraction is lower than expected (X(IGE) $\sim 0.55$ for N1def and N3def models), which is the consequence of the weakly constrained X($^{56}$Ni). Despite the significantly lower densities of the outer regions adopted in our approach, Si, S, Fe and Ni are either absent or present only with a significantly lower mass fraction. \cite{2022MNRAS.509.3580M} argued for strong mixing and uniform abundances, but their work was limited to the analysis of a few lines instead of fitting the whole optical wavelength range and studying a stratification model fixed to the radial coordinates. However, this study, just like previous fitting approaches, does not contradict the results of \cite{2022MNRAS.509.3580M}, that the inner regions may inhibit a nearly uniform abundance structure. 

Moreover, the complete lack or very low abundances of IMEs and IGEs in the outer regions are relatively well constrained in our models by the blue wings of absorption features, and further supported by the low density values at these higher velocities. This result indeed indicates some level of stratification, however, not necessarily a continuously changing multi-layer structure. Instead, a C/O dominated top layer with a deflagration-like well-mixed ejecta is also supported by the spectral synthesis.

\section{Discussion and summary}
\label{sec:discussion_and_summary}

This paper presents a combined analysis of a bright type Iax SN 2018cni and a faint type Iax SN 2020kyg. We attempted to comprehend the nature of the members of type Iax class which are located at the two extremes in the luminosity space of this class. We see that in terms of photometric properties, spectral features, and ejecta configuration these two SNe behave differently. This also reflects that there might be a difference in the progenitor channels for both events. In the following, we summarize our findings for both events and try to make a collective picture of these two events as members of the same class with a gap in luminosity distribution.    

SN 2018cni is a bright member of type Iax class with peak absolute magnitude M$_{V}$ = $-$17.81$\pm$0.21 mag. The estimated bolometric luminosity shows that SN 2018cni share similarity with SN 2005hk and is brighter than SN 2002cx. The mass of $^{56}$Ni estimated from the analytical modeling of the $BgVri$ bolometric light curve is $\sim$ 0.07$\pm$0.01 M${_\odot}$. The spectral features of SN 2018cni match well with SNe 2002cx and 2005hk.

With peak absolute magnitude, M$_{g} = -14.44\pm0.16$ mag, SN 2020kyg lies at the fainter end of type Iax luminosity distribution. It is brighter than SNe 2008ha \citep{stritzinger2014}, 2019gsc \citep{2020ApJ...892L..24S} and 2021fcg \citep{2021ApJ...921L...6K} and fainter than SN 2010ae \citep{stritzinger2014}.  
Analysis of the bolometric light curve ($BgVri$, obtained through \texttt{Superbol}) of SN 2020kyg indicates that $\sim$ 0.002$\pm$0.001 M${_\odot}$ of $^{56}$Ni was produced in the explosion. 

\begin{figure}
	\begin{center}
		\includegraphics[scale=0.35, clip, trim={0.25cm 0.7cm 1.5cm 1.8cm}]{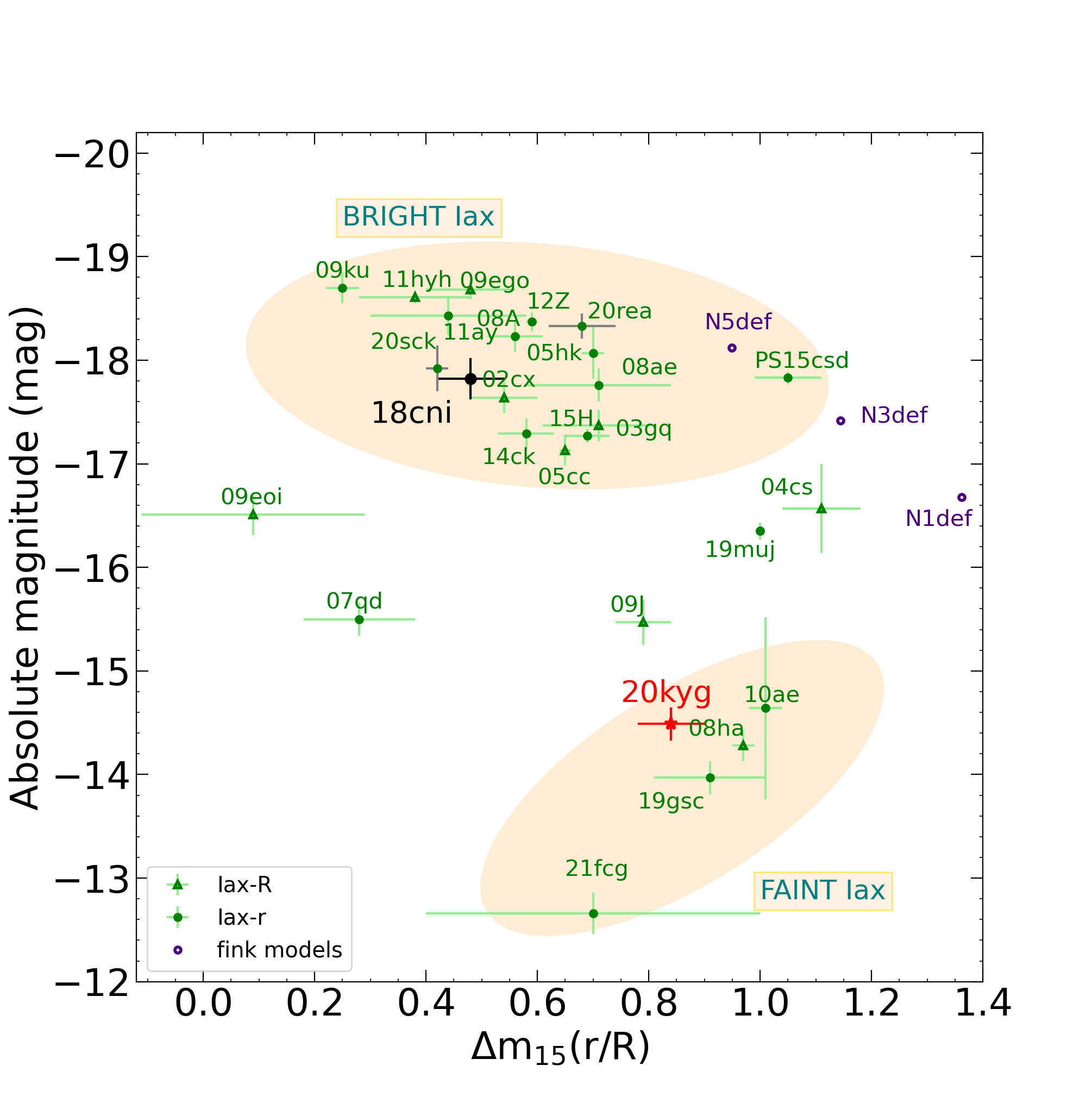}
	\end{center}
	\caption{Absolute magnitude versus light curve decline rate in  $r/R$- band for well studied type Iax SNe.  }
	\label{fig:abs_mag_delta15_2018cni_2020kyg}
\end{figure}

\begin{figure}
	\begin{center}
		\includegraphics[scale=0.3, clip, trim={0.25cm 0.7cm 1.5cm 1.8cm}]{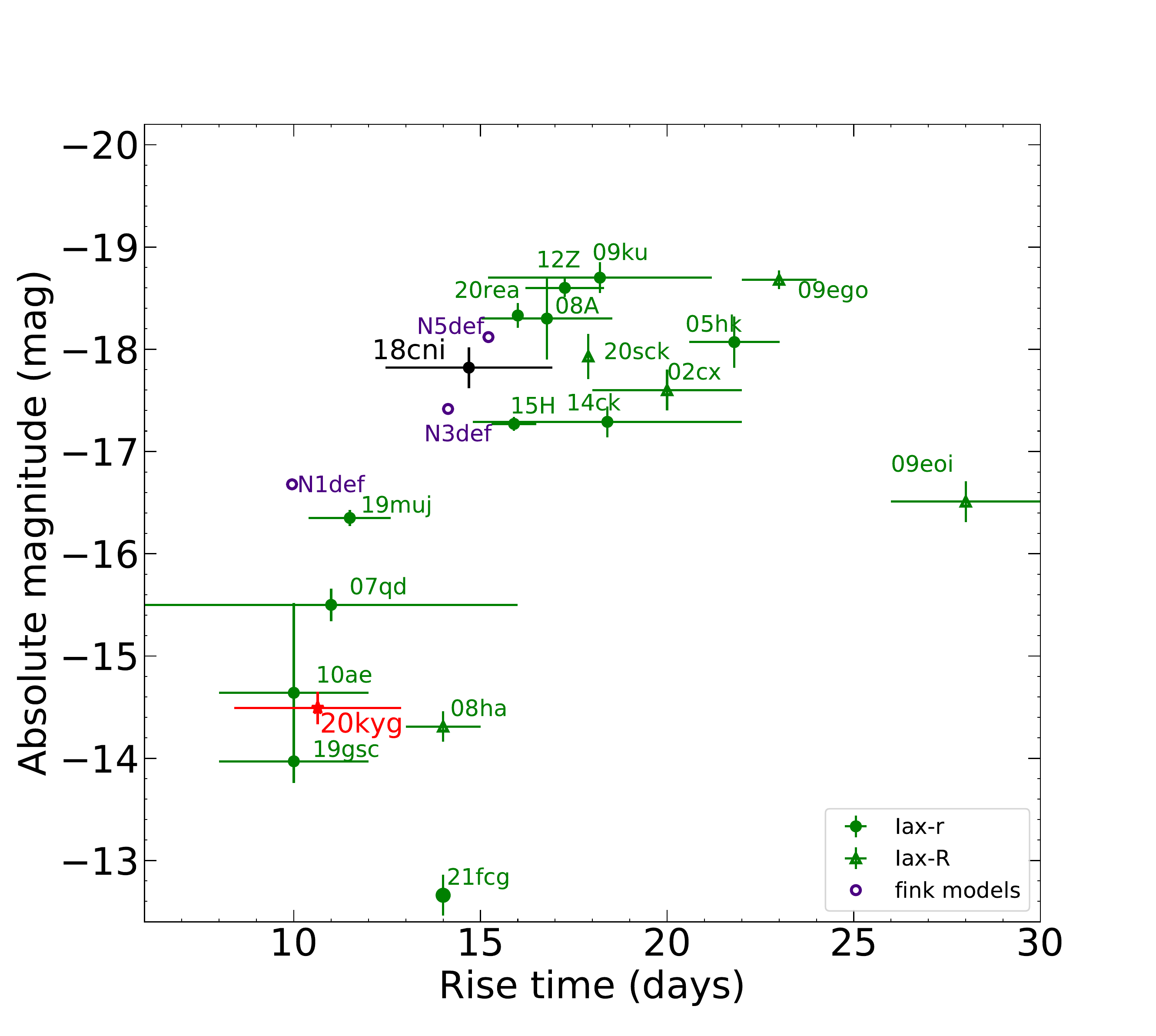}
	\end{center}
	\caption{Distribution of type Iax SNe in terms of the peak absolute magnitude and rise time in $r/R$- band. The figure clearly depicts the fact that type Iax SNe follows a correlation in their peak luminosity and rise time.}
	\label{fig:rise_time_delta15_2018cni_2020kyg}
\end{figure}

\citet{ps15csd2016} suggested a correlation between absolute magnitude, decline rate parameter, and rise time of type Iax SNe in the $R/r$-band. Since then the sample has grown significantly, especially in the fainter end. An attempt is made to check for any possible correlation in the extended sample. Table \ref{tab:ref_fig_17_18} includes all the type Iax SNe we have used in our study and their references. In Figure \ref{fig:abs_mag_delta15_2018cni_2020kyg}, we have plotted the absolute magnitude and decline rate parameter in the $r/R$-band for the extended sample. We have also included N1-def, N3-def, and N5-def models \citep{2014MNRAS.438.1762F} for comparison. We recover the weak correlation between the absolute magnitude and light curve decline rate parameter as reported in \citet{ps15csd2016} and shown in Figure \ref{fig:abs_mag_delta15_2018cni_2020kyg}. However, dividing the current sample based on the brightness of the objects reveals an interesting trend: bright and faint objects behave differently. For bright type Iax cluster (M$_{r}$ $\leq$ $-$17.1 mag), there is a negative correlation with a correlation coefficient of 0.49 (p=0.044), viz. the objects with faster decline rate are found to be fainter. On the other hand, for the faint cluster (M$_{r}$ $\geq$ $-$14.64 mag), a trend of positive correlation emerges (Pearson correlation coefficient of $-$0.85, p = 0.070), viz. objects with faster decline rate appear brighter. The sample of type Iax studied so far is limited and a study with a bigger sample covering both the bright and faint end of luminosity distribution will be required to confirm this trend. In Figure \ref{fig:rise_time_delta15_2018cni_2020kyg}, we have presented a plot between the rise time and absolute magnitude for a sample of well studied type Iax SNe. The rise times of SNe 2018cni and 2020kyg plotted in this figure are estimated from the spectral fitting (Section \ref{sec:Spectral modelling}). From the plot, it is clear that the two quantities are correlated (Pearson correlation coefficient = $-$0.53, p = 0.02), the brighter objects tend to have longer rise time. Type Iax SNe with fair pre-maximum coverage are included in this analysis while some unfiltered measurements are not considered. 

\begin{table}
\caption{References for Figures \ref{fig:abs_mag_delta15_2018cni_2020kyg} and \ref{fig:rise_time_delta15_2018cni_2020kyg}}
\smallskip
\scriptsize
\begin{tabular}{l c}
\hline
SN                         & References         \\
\hline
2002cx, 2003gq, 2004cs    & \cite{2013ApJ...767...57F}  \\
2005cc, 2008ae, 2009J, 2011ay      & \cite{2013ApJ...767...57F}     \\
2005hk     &   \cite{stritzinger2015} \\
2007qd     & \cite{2010ApJ...720..704M}      \\
2008A      & \cite{2012ApJS..200...12H},\\
          & \cite{2013ApJ...767...57F}  \\
2008ha   & \cite{2009AJ....138..376F}, \\
       & \cite{2013ApJ...767...57F}  \\
2009ku   &  \cite{2011ApJ...731L..11N},\\
         &   \cite{2013ApJ...767...57F} \\
2010ae      & \cite{stritzinger2014}    \\
2012Z      & \cite{stritzinger2015}    \\
2014ck    & \cite{2016MNRAS.459.1018T}  \\
PS15csd    &  \cite{ps15csd2016}  \\
PTF09ego, PTF09eoi, PTF11hyh    &  \cite{2015ApJ...799...52W}  \\
2015H   &   \cite{ps15csd2016} \\
2019gsc &   \cite{2020ApJ...892L..24S}, \\
        &    \cite{2020MNRAS.496.1132T} \\
2019muj &   \cite{2021MNRAS.501.1078B}\\
2020sck &   \cite{2022ApJ...925..217D}\\
2020rea &  \cite{2022MNRAS.517.5617S}  \\
2021fcg & \cite{2021ApJ...921L...6K}   \\
2018cni, 2020kyg & this work \\

\hline
\end{tabular}
\label{tab:ref_fig_17_18}      
\end{table}

The spectral features of SNe 2018cni and 2020kyg appear similar, except for the \ion{Si}{2} feature which is prominent in SN 2020kyg. Modeling of the spectral sequence using TARDIS indicates that some level of stratification at the outer layers and mixed ejecta in the inner layers are required.

Comparison of the bolometric light curves obtained through \texttt{Superbol} shows a resemblance of SN 2018cni with N3-def model \citep{2014MNRAS.438.1762F} whereas luminosity of SN 2020kyg falls in the range of pure deflagration of CONe white dwarf. For brighter members of type Iax class, several explosion scenarios have been proposed such as pulsational delayed detonation \citep{1974Ap&SS..31..497I,1991A&A...245L..25K,1991A&A...246..383K,1993A&A...270..223K,1995ApJ...444..831H,1996ApJ...457..500H,2006ApJ...642L.157B,Baron_2012,2014MNRAS.441..532D}, deflagration to detonation transition \citep{1991A&A...245L..25K,1991A&A...245..114K,1993A&A...270..223K,1995ApJ...444..831H,1996ApJ...457..500H,2002ApJ...568..791H,2013MNRAS.429.1156S,2013MNRAS.436..333S}, pure deflagration of white dwarf \citep{2014MNRAS.438.1762F, 2015MNRAS.450.3045K},  core collapse origin \citep{2009AJ....138..376F, 2016MNRAS.461..433F, 2009Natur.459..674V,2010ApJ...719.1445M} etc. Most of the observed properties of relatively bright type Iax SNe match well with the predictions of pure deflagration of white dwarf leaving behind a bound remnant. Observational properties of faint type Iax can be explained by weak deflagration of CONe white dwarf \citep{2015MNRAS.450.3045K}, but the spectro-photometric evolution of these models is faster as compared to the observed ones. Another possible explanation is a double degenerate channel which involves the merger of CO and ONe white dwarfs. This merger results in a failed detonation and is associated with a fast declining transient \citep{2018ApJ...869..140K} fainter than faint type Iax SNe.

In any case, incomplete burning of the white dwarf explains most of the properties of type Iax SNe, but the possibility for another progenitor scenario for faint members of this class cannot be ruled out completely. The different nature of the correlation between peak luminosity and light curve decline rate for bright and faint members indicates heterogeneity within SNe Iax. Further study and in-depth analysis of this class will be required to understand their observed properties, progenitor system, and explosion mechanism. 


\section{Acknowledgments}
{We thank the anonymous referee for providing useful comments towards the improvement of the manuscript. We acknowledge Wiezmann Interactive Supernova data REPository http://wiserep.weizmann.ac.il (WISeREP) \citep{2012PASP..124..668Y}. This research has made use of the CfA Supernova Archive, which is funded in part by the National Science Foundation through grant AST 0907903. This research has made use of the NASA/IPAC Extragalactic Database (NED) which is operated by the Jet Propulsion Laboratory, California Institute of Technology, under contract with the National Aeronautics and Space Administration. This work makes use of the Las Cumbres Observatory global telescope network. The LCO group is supported by NSF grant AST-1911225. RD acknowledges funds by ANID grant FONDECYT Postdoctorado Nº 3220449. DAH, GH, and CM were supported by NSF Grants AST-1313484 and AST-1911225.

This project has been supported by the J\'anos Bolyai Research Scholarship of the Hungarian Academy of Sciences (TS), as well as by the FK134432 grant of the National Research, Development and Innovation Office of Hungary and the \'UNKP 22-5 New National Excellence Programs of the Ministry for Culture and Innovation from the source of the National Research, Development and Innovation Fund (BB \& TS). This research made use of TARDIS, a community-developed software package for spectral synthesis in supernovae \citep{kerzendorf_wolfgang_2018_1292315, kerzendorf_wolfgang_2019_2590539}. The development of TARDIS received support from the Google Summer of Code initiative and from ESA's Summer of Code in Space program. TARDIS makes extensive use of Astropy and PyNE. This work made use of the Heidelberg Supernova Model Archive (HESMA)\footnote{\url{https://hesma.h-its.org}}. We are grateful to the Southern African Large Telescope (SALT) astronomers and support staff who obtained the SALT data presented here. The SALT observations of SN 2018cni were obtained with Rutgers University program 2018-1-MLT-006 (PI: Jha). The Gemini Observatory data for SN 2020kyg were obtained through program GN-2020A-Q-224 (PI: Jha). Observations reported here were obtained at the MMT Observatory, a joint facility of the University of Arizona and the Smithsonian Institution. We would like to express our gratitude to the Lick Observatory staff for their support. We gratefully acknowledge the usage of native lands for our science. Shane 3-m observations were conducted on the stolen land of the Ohlone (Costanoans), Tamyen and Muwekma Ohlone tribes. A major upgrade of the Kast spectrograph on the Shane 3-m telescope at Lick Observatory was made possible through generous gifts from William and Marina Kast as well as the Heising-Simons Foundation. Research at Lick Observatory is partially supported by a generous gift from Google.

\software{TARDIS \citep{kerzendorf_wolfgang_2018_1292315,kerzendorf_wolfgang_2019_2590539}, PYSALT \citep{2010SPIE.7737E..25C}, SuperBol \citep{2018RNAAS...2..230N}, lcogtsnpipe \citep{2016MNRAS.459.3939V}, PyZOGY \citep{2017zndo...1043973G}, floydsspec pipeline (https://www.authorea.com/users/598/articles/6566), spectral pipeline (https://github.com/msiebert1/UCSC)}

\bibliography{ms}{}
\bibliographystyle{aasjournal}



\end{document}